\documentclass[12pt]{article}
\usepackage{a4}
\usepackage{epsf}
\begin{document}
\hspace*{\fill}hep-th/9912071

\begin{center}

{\bf ELECTROWEAK PHASE TRANSITION IN STRONG MAGNETIC
FIELDS IN THE STANDARD MODEL OF ELEMENTARY PARTICLES\\
}
                                ~\\

            {\bf\sl Vladimir Skalozub  and Vadim Demchik\\}                                  ~\\
                                   ~\\

 Dniepropetrovsk State University, 320625 Dniepropetrovsk, Ukraine\\
 e-mail: skalozub@ff.dsu.dp.ua; dvi@ff.dsu.dp.ua
                         ~\\
                         ~\\
                       ABSTRACT
\end{center}

The electroweak phase transition in the magnetic and hypermagnetic
fields is studied in the Standard Model on the base of investigation
of symmetry behaviour within the consistent effective potential of the
scalar and magnetic fields at finite temperature. It includes the
one-loop and daisy diagram contributions. All discovered fundamental
fermions and bosons are taken into consideration with their actual
masses. The Higgs boson mass is chosen to be in the energy interval 75
GeV $\le m_H \le$ 115 GeV.  The effective potential calculated is real
at sufficiently high temperatures due to mutual cancellation of the
imaginary terms entering the one-loop and the daisy diagram
parts. Symmetry behaviour shows that neither the magnetic nor the
hypermagnetic field does not produce the sufficiently strong first
order phase transition.  For the field strengths $H, H_Y$ $\ge
10^{23}$ G the electroweak phase transition is of second order at
all. Therefore, baryogenesis does not survive in the Standard Model in
smooth magnetic fields.  The problems on generation of the fields at
high temperature and their stabilization are also discussed in a
consistent way.  In particular, it is determined that the nonabelian
component of the magnetic field $\ (gH)^{1/2} \sim g^{4/3}T$ has to be
produced spontaneously. To investigate the stability problem the
$W$-boson mass operator in the magnetic field at high temperature is
calculated in one-loop approximation.  The comparison with results
obtained in other approaches is done.

       ~\\
       ~\\

\subsection*{1. INTRODUCTION }

Among interesting problems of nowadays high energy physics there are two ones which, at
first glance, are not connected with each other.  These are the value of the Higgs 
boson mass $m_H$ and the strengths of  magnetic fields
       $H$ which could be present  in the early universe (see surveys
     \cite{She} - \cite{Enq}). Both problems are of paramount
       importance for particle physics and cosmology. For instance, as was shown in
Refs.  \cite{Shap1}, \cite{Elm1}, \cite{Shap4} a large scale
       homogeneous hypercharge magnetic field $H_Y$ must essentially influence
the type of
       the electroweak (EW) phase transition making it strong first order.
 Interest to the effects of strong magnetic fields has considerably increased
recently when it was realized that a standard baryogenesis scenario in  the
Standard Model (SM) of elementary particles  could not be established without   the
fields \cite{Shap2}.   Different mechanisms to produce the fields have been
proposed  \cite{Enol}, \cite{SVZ1}, \cite{Bors2}, \cite{Shapj}, \cite{Enq},
\cite{Baym}-\cite{Enqvist98}, \cite{Oles}. One of the purposes of the present  paper
is  to discuss some recent results on these topics. At the same time, in
 the Monte Carlo simulations and by the nonperturbative methods of quantum field
 theory  \cite{Kal}, \cite{Buch} the magnetic mass of nonabelian gauge fields of
 order $ m_{mag.} \sim g^2 T$  has been determined. It screens the nonabelian
 component of the magnetic field at long distances.  This is  one of the reasons
 why an interest to the hypermagnetic field $H_Y$ was excited. This field, owing
 to its abelian nature,  is not screened at finite temperature. The influence of
 hypermagnetic fields on symmetry behaviour has been first investigated for many  
years ago in Ref. \cite{Kirl} where the similarity to  superconductivity  was
 emphasized.  This important observation plays a decisive role in the description of
the
       EW phase transition. The problem  which requirs further investigations is the
       generation of strong  abelian fields at high temperature. Interesting mechanism
       connecting with an abilian anomaly was suggested in Ref. \cite{Shapg}.

Much more involving is situation with the magnetic field, although it is
  studied for many years. Currently, there are  not common opinions
 about either the ways of producing and stabilization the field or its role at
  high temperature (see papers  \cite{Ska3}, \cite{Enol}, \cite{Fior} and references
  therein). This situation, probably, finds an explanation in the  nonabelian nature
    of the field and lack of simple analogies in condensed matter physics (mainly
       superconductivity is discussed). As an example of  peculiarities, let us
mention the phenomenon  of condensation of the $W$- and $Z$-boson fields which
originates from the
       instability of the perturbative vacuum in strong magnetic fields (see  Refs.
\cite{Ska1} - \cite{AO1}). In such a situation, to treat problems with the
external fields correctly the consistent calculations have to be carried
out. Recent  investigations of the phase transitions in the magnetic fields at high
temperature  \cite{Elmp}, \cite{Pers}, \cite{Fior}  have used as a  qualitative
picture of the phenomenon the description in Refs. \cite{Amol},
       \cite{Enol} derived at a classical level for the case $m_H =
m_Z$ corresponding to second type superconductivity. However,  in these papers  the
effects of fermions (light and heavy) as well as the radiation corrections in  the
fields at high temperature have
    not been included but play an important role. The second goal of the present
       paper is to elaborate the situation with magnetic fields at high
    temperatures.  We shall consider all the problems mentioned within 
    consistent calculations allowing for the one-loop effective potential (EP)
  and the daisy diagram contributions in the external fields at high temperature. To
       find the latter ones the one-loop polarization functions of gauge bosons
 under
  these external conditions will be computed. As it is occured, the influence of
the
   fields crucially depends on the values of the particle masses. So, to have a
reliable quantitative description of the  EW phase transition we fix them to
  be equal to the present day experimental data. The Higgs boson mass will be
taken in the energy interval 75 GeV $\le m_H \le$ 115 GeV. The low bound
corresponds
to the mass values
       when perturbative methods are reliable. The upper bound is chosen to fit
the present experimental low limit $m_H \ge$ 90 GeV. In our calculations, the
external
fields will taken into account exactly through the Green functions. This
insures, the results to be obtained should reproduce correctly the effects
of the   
fields for all the values $m_H$ considered. It is also important to notice that
in strong fields at high temperatures light fermions dominate, as it follows from
the term $H^2 logT/m_f$ entering the one-loop  effective potential, where $m_f$ is
the fermion mass. Actually, at different temperatures the different fermions give 
the dominant contributions.   

The concept of symmetry restoration at high temperature has been intensively
used in studying of the evolution of the universe at its early stages. Nowadays
it
is a corner stone in  investigations of  various problems of cosmology and particle
physics \cite{DK}, \cite{ADL}. In particular, the type of the EW phase
transition and,
        hence, a further evolution of the universe depends on the mass of Higgs
        boson.  As we mentioned above, the idea that strong fields  make the EW
phase  transition strong first order, that is necessary to retain the
standard  baryogenesis (see recent survey \cite{Shap2}), requirs further
investigations with radiation corrections allowed for. Since all the masses of
fundamental  particles, except $m_{H}$, are known one is able to investigate in
detail the phase transition as the function of this parameter  and to determine
the properties of the vacuum. That is the main goal of the  present paper.

Various aspects of the phase transitions in magnetic fields at high temperature
have been discussed in literature \cite{Cha} - \cite{Magp},
        \cite{SVZ2}. In Refs. \cite{Sk1}, \cite{Rez}, considering the boson
part of the Salam-Weinberg model, the EW phase transition  in the strong fields
was investigated and the vacuum structures of the phases  have also been
described. In Ref. \cite{SVZ2} side by side with
temperature
        and magnetic field a chemical potential was incorporated. But the role of
        fermions has not been studied in detail. 
        
Another aspect of the EW phase transition, which also has not been elaborated,
 is the influence of  so-called daisy (or ring) diagrams
  at high temperature and strong fields.  At zero field it has been investigated
in
  Refs. \cite{Tak}, \cite{Din}, \cite{Car} where 
 the importance of these diagrams to correctly describe symmetry behaviour  was
 emphasized. In Ref. \cite{Car} 
  the t-quark mass was chosen of order $110$ GeV.  So, to account of
the experimental value $m_t = 175$ GeV, it has to be revised. In the present
paper the EW
phase transition will be studied for the case of  the
        constant fields $H_Y$ and $H$. This is an adequate approximation for
strong fields in the cases of the second order phase transition and the initial 
stage
of the first order one when the bubbles are not large \cite{Elm1}.

 The content is as follows. In sect. 2, for convenience of readers, the results
of the investigations announced are described in a qualitative manner. In sects. 3, 4  
the contributions of  bosons and fermions to the one-loop EP $V^{(1)}(T, H, \phi_c)$
of classical the scalar $\phi_c$ and the external magnetic fields are
calculated in the form convenient for numeric  investigations. In sect. 5 the
correlation corrections due to daisy
diagrams are computed and  the vacuum stability condition at high temperature is
discussed. Special
        attention is devoted to computation of  the daisy diagrams with the unstable
        (tachyonic) mode presenting in the magnetic field in the $W$-boson spectrum.
        In sect. 6 the restored phase with external fields is described. In sect. 7
the
         high temperature expansion of the EP is present.  Then, in sect. 8
  the EW phase transition in the hypercharge magnetic field is investigated. The same 
 for the magnetic field case is carried out in sect. 9. In these two sections the
detailed analysis of the phase transition is done.
        The one-loop polarization functions of $W$-bosons at high temperatures and
        strong magnetic fields are calculated in sect. 10. This, in particular, 
gives possibility to study self-consistently the vacuum stability. The
spontaneous vacuum  
magnetization at high temperature is investigated in sect. 11. It is 
found that the magnetic fields of order  $(gH)^{1/2} \sim g^{4/3}T$ is generated.
Such strong fields affect all the processes at high temperatures.  
Comparison of the results obtained with that of other approaches is done in
sect. 12.
        Discussion and concluding remarks are given in sect. 13. Appendix
contains
        necessary information on the Mellin transformation technique used in
        calculations of the high temperature asymptotics of the EP. 

\subsection*{2.QUALITATIVE PICTURE OF THE EW  PHASE TRANSITION IN MAGNETIC FIELDS}

As it is belived nowadays, the presence of different kind strong magnetic fields
in the  early universe is rather resonable than exotic phenomenon.  In
literature on this topic
        a lot of dynamic mechanisms  to generate the fields are proposed (see
surveys
        \cite{Enq}, \cite{Oles} and references therein).  In the present paper, we are
   not going to consider all of them. Our goal is to describe a consistent picture of
      the  influence of the fields on the EW phase transition. In this section,
we consider
        in a qualitative manner the mechanisms of producing the external fields,  the
        ways of their action on the vacuum  at zero and finite temperature  and present
        the main results of our investigation. 

\subsection*{2.1 The generation of the primordial magnetic fields}

The generation of the fields can be devided in two classes: 1) generation due to processes
that had happened at the EW phase transition; 2) creation of  the fields before the
EW phase
    transition epoch (for instance, at a GUT scale). To the first class we
refer: 
        the rotation of bubbles at the first order phase transition \cite{Baym};
        fluctuation of gradients of the Higgs field \cite{Vach}; bubble collisions
        \cite{Kibble}, \cite{Ahonen}. These mechanisms  produce microscopic fields
which had to be amplified by magnetohydrodynamic processes up to the macroscopic values
required to fit the astronomic observations, as it is discussed in Refs. \cite{Gail},
\cite{Abrand}. 

To the second class of the processes we refer the spontaneous vacuum magnetization at
        high temperature discovered first at zero temperature by Savvidy
\cite{Sav} and the fields generated by strings \cite{Brand}. Ones had been created,
these
 fields were frozen in a cosmic plasma, evolved with it during the
expansion of the   universe and present at the EW phase transition. 
In what follows, the former mechanism will be investigated
        in detail. We shall show that it does work at high temperature.
So, it could serve to produce the
        seed magnetic field (in contrast to conclusions of  Refs. \cite{Elmp},
        \cite{Pers}).

Let us remind the results of Refs. \cite{Sav}, \cite{SVZ1}, \cite{Enol}, \cite{Ska3}.
The spontaneous vacuum magnetization has been derived from the investigation of the
  EP of covariantly constant (sourceless) chromomagnetic
        field $H^a = H \delta^{a3}$ which is a solution to the classical
Yang-Mills field
equations, where $H =$ const and $a$ is an isotopic index, 
\begin{equation} \label{2.1}
V( H, T ) = \frac{H^2}{2} + V^{(1)}( H, T). 
\end{equation}
It includes the tree-level and the one-loop parts. The minimum of the EP at high 
        temperature $T$ corresponds to the nonzero magnetic field of order
        $(gH)^{(1/2)} \sim g^2 T$, $g$ is gauge coupling constant. 
\footnote{ In paper \cite{Enol} the  important cancellation of logarithmic terms  
entering 
        the zero and the finite temperature parts of the EP was missed. This has been
        resulted in the incorrect value of the vacuum magnetic field spontaneously
created at finite temperature.}
In the EW theory, the $ a = 3$ component of
the  weak isospin just corresponds to the nonabelian part of  usual magnetic
field which we observe in the broken phase. 

Very important for our analysis is the value of the  vacuum field  strength $\sim g^2 T$.
 In fact, as it will be shown in sect. 11, this value is increased when the correlation
 corrections are taken  into acccunt. The field is screened at long distances
        $ l \geq 1/ m_{mag} \sim 1/ g^2 T$ by the magnetic mass of gauge field. 
        However, inside this space domain the strong fields may exist and affect
all the
        processes. Really, typical particle masses are of order $m_T \sim g T$,
        therefore, for small $g $  the Compton wave length $ \lambda_{Compt.} \sim 1/
        m_T$, giving the particle size,  and the Larmor radius, $ r_{Larmor} \sim 1/
        (gH)^{1/2} \sim 1/ g^2 T$ , determining the space range where the charged
        particle spectrum is formed, are both located inside a domain which is  filled
        up by the field. Hence it follows that the field strengths of order $ (gH)^{1/2}
        \sim g^2 T$ or stronger are of interest at high temperature and, in particular,
        the Savvidy mechanism gives rise  such intense fields. 

Interesting mechanism to generate the hypercharge magnetic field due to the abelian
        anomaly was proposed in Ref. \cite{Shapg}. At presen time it is not
        investigated in detail. So, in what follows we will not consider
a consistent picture with this field. We  will just assume that the
strong external field $H_Y$  present when the transition had happened. 

Below, we shall study the  influence of constant the usual magnetic \cite{SVZ2},
        \cite{Bors1} and the hypercharge magnetic \cite{Shap1}, \cite{Elm1},
        \cite{Shap4}, \cite{Bors3} fields. Such the approximation is not
artificial for strong fields. More definitely, the gradients of fields are negligible if
   the relation $\mid \nabla H / H \mid << H / m $ holds \cite{Migd}. Here, $m$
  is a characteristic mass of the problem under consideration. It means that in 
   strong fields the dominat effects are due to intensity  of the fields. At
     high temperature, particle masses are of order $m \sim gT$. As it was pointed
        out in Ref. \cite{Elm1} this approximation works well for the second order
        phase transitions and for the first stages of the first order ones when the
        bubbles are not large. On the other hand, in Refs. \cite{Shapg},
     \cite{Shap1} it was argued that  strong stochastic  hypermagnetic fields are
   able to produce the baryon asymmetry of the universe. These problems are left
        beyond the scope of the present paper.
 
\subsection*{2.2 Mechanisms of acting of the external fields on a vacuum}

The ways in which the magnetic and the hypermagnetic fields affect the vacuum
scalar
   condensate are quite different. In the latter case, it is completely similar to the
    case of superconductivity, as it was investigated first in Ref. \cite{Kirl}. In
        broken phase, the gauge field ( U(1) gauge field in the Higgs model and
        Z-boson field in the EW theory) is screened by its mass. This is the
        consequence of the interaction term $\sim A_{\mu}^2 \phi^2$ presenting in the 
     Lagrangian,  $\phi$ is the scalar field. The influence of the external field is
     reduced to the increase of the vacuum energy and it manifests itself at  tree
    level. In sufficiently strong fields the symmetry restoration happens and the
     gauge field mass $M_A = g \phi_c$ becomes equal zero. For the critical field
     strength one has an estimate, $H^2_Y/2 \sim O (M^4_A)$, which shows that
     the restoration happens when the energy density of the external field
     equals to that of the scalar condensate $\sim M^4_A$. This mechanism
  works both at zero and finite temperature. In the latter case, the critical
value
  of $H_Y$ is decreasing  when the  temperature is increasing, as in  
superconductors. If the mass $m_H \le m_z$, the vacuum of the SM is the first
type "superconductor" with respect to hypermagnetic  field otherwise it is the
second type one. This picture is determined at a classical level \cite{Kirl}.
The external hypermagnetic field delays the first order phase transition making
it stronger, that is necessary for baryogenesis. This conclusion has been
obtain in three approximation in Ref. \cite{Shap1}. As it was discussed in detail
in Ref. \cite{Shap2}, to have a standard baryogenesis scenario the ratio $R =
\Delta \phi_c/T_c$ of the order parameter jump to the critical temperature must
be of order $\sim 1.2 - 1.5 $ At zero external fields, one has $R \sim 0.6$ for
$m_H \sim 70 - 80$ GeV.

In the EW theory, the $U(1)$ symmetry corresponding  to 
 electromagnetic field is retained in the broken phase. Therefore, there are no
couplings of the scalar and the electromagnetic fields
        at tree level. Hence, one would expect that the external fields
affect the vacuum condensate through  the radiation corrections as it was discussed in
Ref. \cite{Sals}. However, in the nonabelian case the influence of  fields on the scalar
condensate is more complicate
and the phase transition  in strong external magnetic fields can happen at tree
level,  because of the non-linearity of the field equations.  The vacuum properties are
 determined by not only the scalar field condensate $\phi_c$ but also the  other 
 order parameter -  so-called $W$-boson condensate \cite{SkaW}, \cite{Sk1},  
\cite{Amol} (see also surveys \cite{Sk2}, \cite{AO1}). Actually, here one faces
        the situation with interacting order parameters. In the EW Lagrangian the next
        two terms, $\sim F_{\mu\nu} W^{+}_{\mu} W^{-}_{\nu}$ and $ \sim
        W^{+}_{\mu} W^{-}_{\mu} \phi_c^2, $  enter, where $F_{\mu\nu}$ is an
        electromagnetic field strength tensor, $W^{\pm}_{\mu}$ is the W-boson field.
        The former interaction is due to a gyromagnetic ratio $\gamma = 2$ inherent the
    nonabelian theories. It is crucial for exciting in strong constant magnetic fields
        $H \geq H_0 = M^2_w/e$ in the $W$-boson spectrum of the tachyonic
(unstable) mode $p_0^2 = p_3^2 + M^2_w - eH$, $p_3$ is a momentum along the
field
direction, which then is condensed owing to the self-interaction $\sim
        (W^{+}_{\mu}W^{-}_{\mu})^2$. The $W$-boson condensate influences  the
scalar field at classical level (due to the second of the above written 
interaction terms). As a result, the
        scalar condensate  is eliminated and the condensates of $W$- and
$Z$-boson 
        fields are formed \cite{Sk2}, \cite{Sk1}, \cite{Amol}. The threshold of the
        phase transition is determined by the value of the Higgs boson mass. If $m_H
        < M_w$ the vacuum is the first type ``supercondustor'' with respect to
the magnetic field. In this case  the homogeneous 
         $W$-condensate is
        formed  \cite{SkaW}. For $m_H > M_w$ the vacuum behaves as  a second
        type "superconductor", therefore the lattice structure of the Abrikosov
type formed by the W- and Z-boson fields is
        produced. This picture has been derived for arbitrary mass $m_H$ in Refs.
        \cite{Sk1}, \cite{MacD} and for the special value $m_H = m_z$ in Refs.
        \cite{Amol}, \cite{AO1}. 

These vacuum properties have been determined in tree approximation when
fermions  do not contribute. But they  do contribute in
 one-loop order, and, moreover, play  a decisive role in the vacuum dynamics. 
In strong fields at high temperatures the influence of light fermions is increased, as 
the consequence of the term $\sim H^2 log(T/m_f)$ entering the one-loop EP. Besides,
 other peculiarities must be taken into consideration. Namely, the EP contains
additional
 $T$-dependent terms, such as $\sim \phi^2_c T^2, \sim eH T \phi_c$, etc. (see
 more details in sects. 5, 7), which are generated in the one-loop and higher
orders   of perturbation theory. These terms make the picture of the field
action involving.
 Therefore, to investigate symmetry behaviour numeric calculations must be
carried out. 

\subsection*{ 2.3 Symmetry behaviour in magnetic fields }

 As it is well known \cite{Tak}, \cite{Car}, \cite{Din}, at finite temperature
side
by side with
 the one-loop EP the correlation corrections described by the daisy diagrams
  should be  added in the consistent calculations. Series of these diagrams are
    responsible for the long range effects and contain imaginary terms which
   cancel  the imaginary part of the one-loop EP. As a result, the total EP is real at
        sufficiently high temperature. This important property is fulfilled also
in the external fields when the contribution of daisy diagrams with unstable mode
is allowed for \cite{Bors1}, \cite{Bors2}.

Now, we are going to describe  symmetry behaviour at high temperature 
and strong magnetic fields.  In what follows we assume that only one type of the 
fields is applied. Remind that in the broken phase the component of  $H_Y$
responsible for the  
magnetic field is
       unscreened and the form of the EP curve corresponds to both the magnetic
and  the hypemagnetic fields.  In  order to investigate the EW phase transition we
       shall consider the function $ {\cal V}^{'} = {\cal V}(h,B,\phi) - {\cal
       V}(h, B, 0) $ describing the symmetry restoration. Here, the dimensionless  
magnetic, $h = H/H_0, H_0 = M^2_w/e,$ and the scalar, $\phi = \phi_c/\delta(0)$,
fields and the
inverse temperature, $B = \beta M_w$, are introduced, $\delta(0)$ is the EP
minimum  
position at zero temperatures and fields (more details see in sects. 8, 9).
The normalized order parameter $\phi$ is changed from unit to zero that is
convenient in numeric calculations.

In fact, symmetry breaking (restoration) can be realized in two ways. 
By the first order phase transition with a nonzero jump of the order parameter 
$\Delta \phi_c \ne 0$ or by the second order one when the order parameter
 is changing smoothly.

 In Fig. 1 we depict  symmetry behaviour at high temperatures for the first order phase
transition. Notice that we consider  the  metastable electroweak minimum  which
is  
separated from
a global unbounded state disposed at large values of $\phi^2$ by a wide and high  
potential
barrier. The temperatures $T_{c_{1}}$ and $T_{c_{2}}$ are called the spinodal  
temperatures, 
which correspond  the situations when bubles of the broken $(T_{c_{1}})$ and the restored
$(T_{c_{2}})$ phases can not exist in the vacuum.

\begin{figure}
\begin{center}
  \epsfxsize=0.6637222\textwidth
  \epsfbox[0 100 600 500]{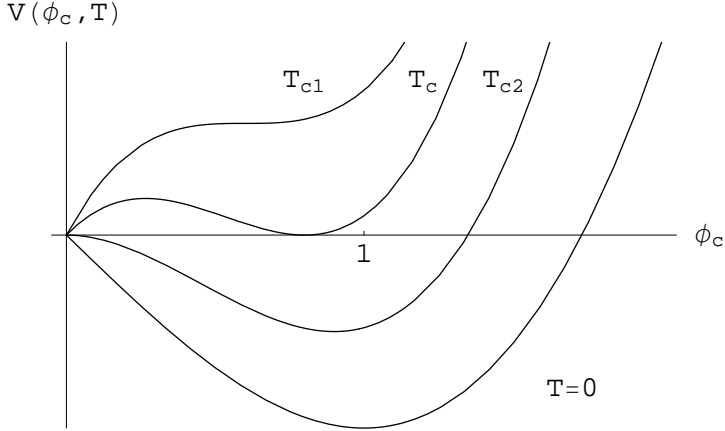}
  \caption{  Symmetry behaviour for the first order phase transition.
$T_{c_1}$ and
$T_{c_2}$ are upper and lower spinodal temperatures, $T_c$ is critiacal temperature.}
\end{center}
\end{figure}
\begin{figure}
\begin{center}
  \epsfxsize=0.6637222\textwidth
  \epsfbox[0 100 600 500]{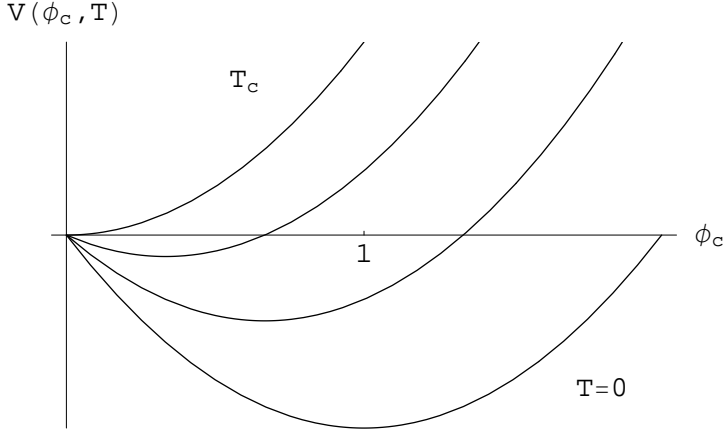}
  \caption{  Symmetry behaviour for the second order phase transition. The
order
parameter $\phi_c(H, T)$  is increasing  monotonically with temperature decreasing.
               }\end{center}
\end{figure}

In Fig. 2 we show symmetry behaviour for the second order phase transition. In this 
case baryogenesis can not be realized because of washing away the  
baryon-antibaryon asymmetry (see survey \cite{Shap2}).

As we mentioned before, at zero external fields the value of the parameter  $R
\sim
0.6$ and so
baryogenesis is not possible in the SM. Our main task here is to investigate  the
situation with strong external fields been taken into consideration.  

The general picture of the phase transitions is independent of the specific
external conditions considered. The external fields change the parameters (or the
type) of
 the phase transitions. Besides, because of the different mechanisms of influence of the
magnetic and hypermagnetic fields on the vacuum the conditions fixing the transition
temperature are also different for these fields (for more details see sects. 8, 9).

Now, let us describe the restored phase. For the magnetic
field case, because of the presence of the tachyonic mode in the W-boson spectrum,
the state $\phi = 0$ is unstable
at zero temperature. To better understand the   situation at $T \ne 0$, let us first
discussed the
 one-loop case. To verify whether the vacuum is stable or not one has to consider 
the effective mass squared of the unstable mode defined as
 the sum of the ground state energy squared taken at zero momentum and the
one-loop W-boson mass operator averaged  over the ground state. It must be
considered
for the value of $H$ condensed at high temperature. In particular,  we have for
the field
$H^{(1)}$ condensed at the one-loop level  $ (gH)^{1/2}_c = (g^2/2\pi) T  
\cite{Bors1}$: $ M^2(H_c, T) = \Pi(H_c, T, n=0, \sigma
  =+1) - gH_c  > 0 $. Thus, the vacuum stabilization is observed.
The situation which takes place when the correlation corrections are included  is
discussed in sect. 11.

 In the restored phase the $W$-bosons do not interact  with the hypermagnetic field.
 So, no instabilities occur in this case. 

By investigating symmetry behaviour in either the hypermagnetic or the magnetic field
within the total EP inlcuding the contributions of all the SM particles we have
determined  by numeric computations that  for all $m_H$ values  considered the
increase in the field strength is resulted in the weaker (not stronger) first order phase
transition. For $H, H_Y \geq 0.1 - 0.5~~10^{24}$ G it becomes of the second order at 
all. Thus, we come to the conclusion that  baryogenesis does not survive at
these external conditions.

 To better understand the role of
fermions in symmetry behaviour let us adduce two terms of the
asymptotic expansion of the one-loop EP in the limit of $T \rightarrow \infty, H
\rightarrow
\infty$. The first one is the term
$\sim H^2 log \frac{T}{m_f}$. Due to this term the light fermions are
dominant at high temperature.
The second term can be derived from the expansion of the zero temperature part:
$\sim - eH m_f^2 $
 This term acts to make "heavier" the Higgs particles
in the field.  As a result, the second order temperature phase transition  is
stimulated due to strong fields.

We would like to complete this section with the comparison of the
described results with that of other approaches (for more details see sect. 8, 12).
The EW phase transition in the hypermagnetic field has been investigated in one-loop
approximation to the EP in Ref. \cite{Elm1} and by the method
combining perturbative results and lattice simulations in Ref. \cite{Shap4}. These  
authors bacause of different reasons have skipped the fermion part of the EP
 and therefore had no  possibility to determine the form of the
EP curve in strong fields at high temperature. In this respect our investigation
filled up the gap existed.
  In our investigation the external fields have been taken into account exactly
through Green's functions. Therefore, in particular, influence of fermions on symmetry
behaviour was correctly reproduced.

\subsection*{3. BOSON FIELD CONTRIBUTIONS TO $V^{(1)}(T,H,\phi_c)$}

The Lagrangian of the boson sector of the Salam-Weinberg model is (see, for example,
        \cite{Chen})

\begin{eqnarray} \label{1}
 L  = -\frac{1}{4} F^a_{\mu\nu} F^{\mu\nu}_a
-\frac{1}{4} G_{\mu\nu} G^{\mu\nu} + (D_{\mu}\Phi)^+ (D^{\mu}\Phi)
\nonumber\\
+\frac{m^2}{2}(\Phi^+ \Phi) - \frac{\lambda}{4} (\Phi^+\Phi)^2,
\end{eqnarray}
where the standard notations are introduced
\begin{eqnarray}
F^a_{\mu\nu} = \partial_{\mu}A^a_{\nu}-\partial_{\nu}A^a_{\mu}
+ g\varepsilon^{abc} A^b_{\mu}A^c_{\nu},
\nonumber\\
G_{\mu\nu}  = \partial_{\mu}B_{\nu}-\partial_{\nu}B_{\mu},
\nonumber\\
D_{\mu}  =  \partial_{\mu} + \frac{1}{2}ig A^a_{\mu}\tau^a + \frac{1}{2}ig'
B_{\mu}.
\nonumber\\
\end{eqnarray}
The vacuum expectation value of the field $\Phi$ is
\begin{equation} \label{2} 
<\Phi> = \left(\begin{array}
{c}0 \\ \phi_c \end{array} \right).
\end{equation}
The fields corresponding to the $W$-, $Z$-bosons and photons, respectively, are
\begin{eqnarray}  
W^{\pm}_{\mu} =\frac{1}{\sqrt{2}}(A^1_{\mu} \pm iA^2_{\mu}),
\nonumber\\
 Z_{\mu} =\frac{1}{\sqrt{g^2 + g'^2}}(gA^3_{\mu} - g'B_{\mu}),
\nonumber\\
 A_{\mu} =\frac{1}{\sqrt{g^2 + g'^2}}(g'A^3_{\mu} + gB_{\mu}).
\nonumber\\
\end{eqnarray}

To incorporate interaction with an external hypermagnetic field we add the term
$\frac{1}{2}
        \vec{H}\vec{H_Y}$ to the Lagrangian. The value of the macroscopic magnetic
        field generated inside the system will be determined by minimization of free  
energy.  Interaction with classical electromagnetic field is introduced as usually by
splitting the
        potential in
two parts: $A_{\mu}= \bar{A_{\mu}} + A^{R}_{\mu} $, where $A^{R}$ describes a
radiation field and $\bar{A} = (0,0,Hx^1,0)$ corresponds to the constant
magnetic field directed along the third axis. We make use of the gauge-fixing
conditions \cite{Sk2}
\begin{equation} \label{3} 
\partial_{\mu}W^{\pm \mu} \pm ie\bar{A_{\mu}}
W^{\pm \mu} \mp i\frac{g\phi_c}{2\xi}\phi^{\pm} = C^{\pm}(x),
\end{equation}
\begin{equation} \label{4} \partial_{\mu}Z^{\mu} - \frac{i}{\xi'}
(g^2 + g'^2)^{1/2}\phi_c \phi_{z} = C_z(x) ,
\end{equation}
where $ e = g sin \theta_w, tan \theta_w = g'/g, \phi^{\pm}, \phi_{z}$ are the
Goldstone fields, $\xi, \xi' $ are the
gauge fixing parameters, $C^{\pm}, C_z$ are arbitrary functions and $\phi_c$ is
a value of the scalar field condensate. In what follows, all calculations will be
carried out in
the general relativistic renormalizable gauge (\ref{3}),(\ref{4}) and after
that we  set $\xi,\xi' = 0$ choosing the unitary gauge.

To compute the EP $V^{(1)}$ in the background magnetic field let us
introduce the proper time, s-representation, for the Green functions
\begin{equation} 
G^{ab}= - i \int\limits_{0}^{\infty} ds \exp(-is {G^{-1}}^{ab})
\nonumber\\
\end{equation}
and make use the
method of Ref. \cite{Cab}, allowing in a natural way to incorporate the temperature
into this formalism. A basic formula of Ref. \cite{Cab} connecting the Matsubara
Green functions with the Green functions at zero temperature is needed,
\begin{equation} \label{5} 
G^{ab}_k(x,x';T) = \sum\limits_{-\infty}^{+\infty}
(-1)^{(n+[x])\sigma_k} G^{ab}_k(x-[x]\beta u, x'- n\beta u),
\end{equation}
where $G^{ab}_k~ $is the corresponding function at $T=0, \beta =1/T, u =
(0,0,0,1),$ the symbol $[x]$ denotes an integer part of $x_{4}/\beta, \sigma_k
= 1$
in the case of physical fermions and $\sigma_{k} =0$ for boson and ghost fields.
The Green functions in the right-hand side of formula (\ref{5}) are the matrix
elements of the operators $G_k$ computed in the states $\mid x',a)$ at $T=0$,
and in the left-hand side the operators are averaged over the states with $T\not=
 0$. The corresponding functional spaces $U^{0}$ and $U^{T}$ are different but
in the limit of $T \to 0$ $ U^{T}$ transforms into $U^{0}$.

The one-loop contribution into EP is given by the expression
\begin{equation} \label{6} V^{(1)} = - \frac{1}{2} Tr\log G^{ab},
\end{equation}
where $G^{ab}$ stands for the propagators of all the quantum fields $W^{\pm},
\phi^{\pm},...$ in the background magnetic field $H$. In the s-representation
the calculation of the trace can be done in accordance with formula \cite{Sch}
\begin{equation} 
Tr\log G^{ab} = - \int\limits_{0}^{\infty} \frac{ds}{s}
tr \exp(-is G^{-1}_{ab} ).
\nonumber
\end{equation}
Details of calculations based on the s-representation and the formula (\ref{5})
can be found in Refs. \cite{Cab}, \cite{Rez}, \cite{Ska3}. 
The terms with $n=0$ in Eqs.(\ref{5}),
(\ref{6}) give  zero temperature expressions for  the Green functions and the
effective potential $V^{(1)}$, respectively. They are the only terms possessing
divergences. To eliminate them and uniquely fix the potential we use the
following renormalization conditions at $H, T = 0$ \cite{Rez}:
\begin{equation} \label{7} 
\frac{\partial^2 V(\phi,H)}{\partial H^2}\mid_{H=0,
\phi=\delta(0)} = \frac{1}{2},
\end{equation}
\begin{equation} \label{8} 
\frac{\partial V(\phi,H)}{\partial \phi}\mid_{H=0,
\phi=\delta(0)} = 0,
\end{equation}
\begin{equation} \label{9} \frac{\partial^2 V(\phi,H)}{\partial \phi^2}
\mid_{H=0,\phi=\delta(0)} = \mid m^2 \mid,
\end{equation}
where $V(\phi,H)=V^{(0)}+V^{(1)}+ \cdots$ is the expansion in a number of loops
and $\delta(0)$ is the vacuum value of the scalar field determined in   tree
approximation.

It is convenient for what follows to introduce the dimensionless quantities:
$h=H/H_0 (H_0=M^2_w/e),\phi=\phi_c/\delta(0), K =m_H^2/M_w^2,$ $B=\beta M_w,
\tau=1/B = T/M_w,$ $ {\cal V}= V/H^2_0$ and $M_w = \frac{g}{2}\delta(0)$. 
After reparametrization the scalar field potential is explicitly expressed
through the ratio $K,$
\begin{equation} \label{10} 
{\cal V}^{(0)} = \frac{h^2}{2} + K\sin^2 \theta_w(-\phi^2
 + \frac{\phi^4}{2} ).
\end{equation}
Notice that $h$  in the case of the external hypermagnetic field  is the  component
of $h_Y$ which remains unscreened in the broken
        phase. In the restored phase, it will be convenient to work in terms of the
initial
        fields and we will carry out the corresponding calculations later. 

The renormalized one-loop EP is given by the sum of the functions
\begin{equation} \label{11}
{\cal V}_1 = {\cal V}^{(0)} + {\cal V}^{(1)}(\phi,h,K)
 + \omega^{(1)}(\phi,h,K,\tau),
\end{equation}
where ${\cal V}^{(1)}$ is the one-loop EP at $T=0$, which has been studied
already in Ref. \cite{Sk2}. It has the form:
\begin{equation} \label{12} {\cal V}^{(1)} = {\cal V}^{(1)}_{w,z} +
{\cal V}^{(1)}_{\phi},
\end{equation}
where
\begin{eqnarray} \label{12a} {\cal V}^{(1)}_{w,z} = &\frac{3\alpha}{\pi}&
[h^2 log \Gamma_1 (\frac{1}{2} +\frac{\phi^2}{2h}) + h^2 \zeta^{'}(-1) +
\frac{1}{16} \phi^4 - \frac{1}{8} \phi^4 log\frac{\phi^2}{2h} + \frac{1}{24}
h^2 \nonumber\\  &-& \frac{1}{24} h^2 log(2h)]
+ \frac{\alpha}{2\pi} [ - 2h^2 + (h^2 + h \phi^2) log(h + \phi^2)\nonumber\\ 
&+& (h^2 - h\phi^2 ) log \mid h - \phi^2 \mid ]
+ i \frac{1}{2} \alpha h (\phi^2 - h) \theta (h - \phi^2),
\end{eqnarray}

\begin{eqnarray} \label{12b} {\cal V}^{(1)}_{\phi} &=&
+  \frac{3 \alpha}{4\pi}( 1 + \frac{1}{2 cos^4 \theta_w}) (\frac{1}{2} \phi^4
log \phi^2 - \frac{3}{4} \phi^4 + \phi^2 )  \nonumber\\
&+& \frac{\alpha K^2}{32 \pi} [(\frac{9}{2} \phi^4 - 3 \phi^2 +
\frac{1}{2} ) log \mid \frac{3 \phi^2 - 1}{2} \mid - \frac{27}{4}\phi^4 +
\frac{21}{2} \phi^2 ]
\end{eqnarray}
and $\omega^{(1)}$ is the temperature dependent contribution to the EP
given by the terms of formulae (\ref{5}), (\ref{6}) with $n \not= 0$.

We outline the used calculation procedure considering the $W$-boson
contribution as an example \cite{Ska3},
\begin{eqnarray} \label {13}
\omega^{(1)}_{w} =  \frac{\alpha}{2\pi}
\int\limits_{0}^{\infty}\ \frac{ds}{s^2}\ e^{-is(\phi^2/h)} \Bigl[\frac{1 +
2 \cos 2s}{\sin s} \Bigr]
\sum\limits_{1}^{\infty} \exp(ihB^2 n^2/4s).
\end{eqnarray}
As Eq. (\ref{12a}), this expression contains an imaginary part for $h > \phi^2$
appearing due to the tachyonic mode $ \varepsilon^2 = p^2_3 + M^2_w - eH $ in
the $W$-boson spectrum \cite{Sk2}. It can be explicitly calculated by means of the
analytic continuation taking into account the shift $s \rightarrow s -i0 $ in the $s$-plane. 
Fixing $\phi^2/h > 1$ one can rotate clockwise the integration contour
in the $s$-plane and direct it along the negative imaginary axis. Then, using
the identity
\begin{equation} \frac{1}{\sinh s} = 2 \sum\limits_{p=0}^{\infty} e^{-s(2p+1)}
\end{equation}
and integrating over $s$ in accordance with the standard formula
\begin{equation} \label{14} \int\limits_{0}^{\infty} ds s^{n-1} \exp(-\frac{b}
{s} - as) = 2 (\frac{b}{a})^{n/2} K_{n}(2\sqrt{ab}),
\end{equation}
$a,b > 0$, one can represent the expression (\ref{13}) in the form
\begin{equation} \label{14a}  Re \omega^{(1)}_w = - 4 \frac{\alpha}{\pi}
\frac{h}{B} ( 3\omega_0 + \omega_1 - \omega_2 ),
\end{equation}
where
\begin{equation} \omega_0 = \sum\limits_{p=0}^{\infty} \sum\limits_{n=1}^
{\infty} \frac{x_p}{n} K_1(nBx_p) ; x_p = (\phi^2 + h +2ph )^{1/2}
\end{equation}
\begin{equation} \omega_1 = \sum\limits_{n=1}^{\infty} \frac{y}{n} K_1(nBy),
y = (\phi^2 - h )^{1/2}
\end{equation}
and in the range of parameters $ \phi^2 < h $ after analytic continuation
\begin{equation} \omega_1 = -\frac{\pi}{2} \sum\limits_{n=1}^{\infty}
\frac{\mid y \mid}{n} Y_1(nB\mid y \mid) ,
\end{equation}
\nonumber\\
\begin{equation} 
\omega_2 = \sum\limits_{n=1}^{\infty} \frac{z}{n} K_1(nBz),
z = (\phi^2 + h )^{1/2},
\end {equation}
 $K_n(x), Y_n(x)$ are the MacDonald and Bessel functions, respectively. The
imaginary part of $\omega^{(1)}_w$ is given by the expression
\begin{equation} \label{15} 
Im \omega_1= -2\alpha\frac{h}{B}\sum\limits_{n=1}
^{\infty} \frac{\mid y \mid }{n} J_1(nB\mid y \mid),
\end{equation}
$J_1(x) $ is Bessel function. As it is well known, the imaginary term of the EP is
signaling the instability of a system. In what follows, we shall
consider mainly symmetry behaviour described by the real part of the EP.
As  the imaginary part is concerned,  it will be cancelled  in  a consistent
calculation 
including the one-loop and daisy diagram contributions to the EP.

Carrying out similar calculations for the $Z$- and Higgs bosons, we obtain
\cite{Rez}:
\begin{equation} \label{16} \omega_z = - 6\frac{\alpha}{\pi} \sum\limits_{n=
1}^{\infty} \frac{\phi^2}{\cos^2 \theta_w n^2 B^2} K_2(\frac{nB\phi}{\cos\theta})
\end{equation}
\begin{equation} \label{17}
Re \omega_{\phi} = \Bigl\{
\begin{array}{c}-2 \frac{\alpha}{\pi} \sum\frac{t^2}{B^2 n^2} K_2(nBt)\\
\alpha \sum\limits_{n=1}^{\infty} \frac{\mid t \mid^2}{n^2 B^2} Y_2(nB\mid t
\mid)\end{array}\Bigr\}.
\end{equation}
where the variable $t = [K_w(\frac{3\phi^2-1)}{2})]^{1/2}$ at $3\phi^2>1$ and
the series with the function $Y_2(x)$ has to be calculated  at $3\phi^2<1$.
The corresponding imaginary terms are also cancelled as it will  be  shown below.

The above expressions (\ref{12}), (\ref{14a}), (\ref{16}), (\ref{17}) will be
used in numerical studying of symmetry behaviour at different $H, T$. 
Notice the cancellation of a number of terms entering the zero-temperature
part given Eqs. (\ref{12}) and $T$-depended one. This fact has a
general character and was used in checking of the correctness of calculations.

\subsection*{4. Fermion contributions to $V^{(1)}(H,T,\phi_c)$}

The fermion one-loop EP in magnetic fields is well studied \cite{Sch}, \cite{Ditt1}, 
\cite{Elm}. To find this explicit form at finite temperature let us consider
 the standard  unrenormalized  expression written in the $s$-representation

\begin{equation} \label{18} V^{(1)}_{f} = \frac{1}{8\pi^2}\sum\limits_
{n=-\infty}^{\infty} (-1)^n \int\limits_{0}^{+\infty}\frac{ds}{s^3}
e^{-(m^2_fs+\beta^2n^2/4s)} (eHs)coth (eHs) ,
\end{equation}
$m_f$ is the fermion mass. Here, we have incorporated the equation (\ref{5}) to
include a temperature dependence. In what follows, we shall allow for
the contributions of all fermions - leptons and quarks - with their present masses.
 Usually, considering  symmetry behaviour without field
one restricts himself by a $t$-quark  contribution, only. But in the case of
external fields applied this is not a good idea, since the dependence of $V^{(1)
}$ on $H$ is a complicate function of the ratio $m^2_{f}/eH$. So, at some
fixed values of $H, T$ the different kind dependencies on $H$ will contribute for
fermions with different masses. Hence, a very complicate dependence on the field takes 
place in general. We shall include this in a total by carrying out  numerical
computetions and summing up over all the fermions. Now, separating a zero
temperature contribution by  means of the relation $\sum\limits_{-\infty}^
{+\infty} = 1 + 2 \sum\limits_{1}^{\infty}$ and introducing the parameter
$K_{f}=m^2_{f}/M^2_w = G^2_{Yukawa}/g^2$, we obtain for the  zero temperature
fermion contribution to the dimensionless EP,
\begin{eqnarray} \label{19}
{\cal V}_{f}(h,\phi) = &\frac{\alpha}{4\pi}& \sum\limits_{f} K^2_{f}(- 2 \phi^2
+ \frac{3}{2} \phi^4 - \phi^4 log \phi^2 )\nonumber\\
 &-&\frac{\alpha}{\pi}\sum\limits_{f}(q^2_f\frac{h^2}{6}
\log\frac{2\mid q_f \mid h}{K_f})
\nonumber\\
&-& \frac{\alpha}{\pi} \sum\limits_{f} \Bigl.[2q^2_f h^2 \log\Gamma_1(
\frac{K_f\phi^2}{2\mid q_f\mid h}) + (2\zeta'(-1)-\frac{1}{6})q^2_f h^2
\nonumber\\
&+&\frac{1}{8} K^2_f\phi^4 + (\frac{1}{4}K^2_f\phi^4 - \frac{1}{2}K_f\mid q_f
\mid h\phi^2) \log\frac{2\mid q_f \mid h}{K_f \phi^2}\Bigr],
\end{eqnarray}
where $q_f$ is a fermion electric charge, the sum $\sum\limits_{f} = \sum
\limits_{f=1}^{3}(leptons) + 3 \sum\limits_{f=1}^{3}(quarks)$ counts the
contributions of leptons and quarks with their electric charges.  The $\Gamma_1$
function is defined as follows (see Refs. \cite{Ditt1}, \cite{Sk2}):
\begin{equation} \log\Gamma_1(x) = \int\limits_{0}^{x} dy \log\Gamma(y) +
\frac{1}{2}x(x-1) - \frac{1}{2}x\log(2\pi).
\end{equation}
\nonumber\\

The finite temperature part can be calculated in a way described in
the previous section. In the dimensionless variables it looks as follows:
\begin{eqnarray} \label{20} \omega_{f}&=& 4 \frac{\alpha}{\pi}\sum
\limits_{f}\Bigl\{\sum_{p=0}^{\infty}\sum_{n=1}^{\infty}(-1)^n \Bigl[\frac{
(2ph + K_f\phi^2)^{1/2} h}{Bn} K_1((2ph + K_f\phi^2)^{1/2}Bn)
\nonumber\\
&+& \frac{((2p+2)h + K_f\phi^2)^{1/2}}{Bn} h
K_1(((2p+2)h + K_f\phi^2)^{1/2}Bn)\Bigr]\Bigr\}.
\end{eqnarray}
Again, a number of terms in Eqs. (\ref{19}) and (\ref{20}) are cancelled in the
total, as in the bosonic sector.

These two expressions will also be used in numeric investigations of symmetry behaviour. 

\subsection*{5. CONTRIBUTION OF DAISY DIAGRAMS}

It was shown by Carrington \cite{Car} that at $T \not = 0$ the consistent calculation 
of the EP based on the generalized propagators, which include the polarization
operator insertions, requires that daisy (ring) diagrams have to be added simultaneously
 with the one-loop contributions. These diagrams  essentially affect
the phase transition at high temperature and zero field \cite{Tak}, \cite{Car},
\cite{Din}.
 Their importance at $ T$ and $H \not= 0$ was also pointed out in Refs. 
 \cite{Bors1}, \cite{Bors2}. 

As it is known \cite{Kal}, \cite{Tak}, the sum of daisy diagrams describes a dominant
contribution of long distances. It is important when massless states appear in a system.
 So, this type of diagrams has to be allowed for when a
symmetry restoration is investigated. To find the correction $V_{ring}(H,T)$ at
high temperature and magnetic field the polarization operators of the Higgs
particle, photon and $Z$-boson at the considered background should be
calculated. Then, $V_{ring}(H,T)$ is given by a series depicted in figures 3, 4.
\begin{figure}
\begin{center}
  \epsfxsize=0.6637222\textwidth
  \epsfbox[0 250 612 352]{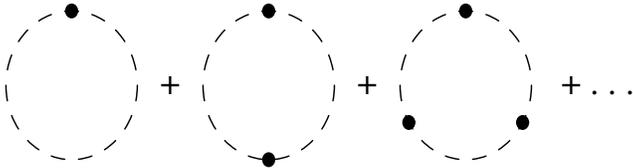}
  \caption{The Higgs field daisy diagrams giving contribution to the effective
potential. Blobs stand for the neutral scalar field polarization operator calculated
at zero momentum.} 
\end{center}
\end{figure}
\begin{figure}
\begin{center}
  \epsfxsize=0.6637222\textwidth
  \epsfbox[0 250 612 352]{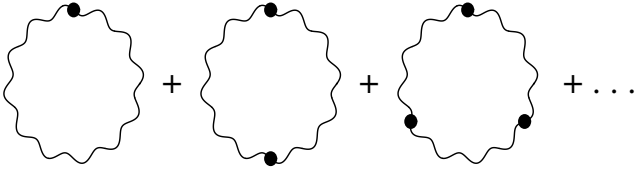}
  \caption{The photon and Z-boson daisy diagrams giving contribution to the
effective
potential. Blobs stand for the polarization operators of the fields calculated at
zero
momenta. }
\end{center}
\end{figure}
Here, a dashed line describes the Higgs particles, the wavy lines represent photons
and $Z$-bosons, the blobs correspond to the polarization operators in the limit of
zero momenta. As also it is known \cite{Tak}, \cite{Car}, in order to calculate the
contribution 
of daisy diagrams   the limiting expressions of the polarization operators $\Pi_{\mu\nu}(k,T,H)$
 at zero momenta, $\Pi_{00}(k=0,T,H)$, are to be substituted. This limit, called the
Debye mass, can be computed  from the EP of
the special type. The latter fact considerably simplifies our task.

Now, let us turn to calculations of $V_{ring}(H,T)$. It is given by the standard expression
 \cite{Kal}, \cite{Tak}, \cite{Car}, \cite{SVZ1}:
\begin{equation} \label{21}
V_{ring} = - \frac{1}{12\pi\beta} Tr\{[M^2(\phi) +
\Pi_{00}(0)]^{3/2} - M^3(\phi)\},
\end{equation}
where trace means the summation over all the contributing states, $M(\phi)$ is
the tree mass of the corresponding state. The functions  $\Pi_{00}(0)$ are:
 $\Pi_{00}(0)$ =$ \Pi(k=0,T,H)$ for the
Higgs particle; $\Pi_{00}(0)=\Pi_{00}(k=0,T,H)$ - the zero-zero
components
of the polarization functions of gauge fields in the magnetic field taken at
zero
momenta.
The above contributions are of order $ \sim g^3 (\lambda^{3/2}) $ in the
coupling  
constants whereas the two-loop terms have order $\sim g^4
(\lambda^{2})$. For
$\Pi_{00}(0)$ the high temperature limits of polarization functions have
to be substituted which are of  order $\sim T^2$ for leading terms and
$\sim g\phi_c T, (gH)^{1/2}T (\phi_c/T << 1 , (gH)^{1/2}/T << 1)$ for subleading
ones.

\subsection*{ 5.1 The polarization function of scalar field}

For the next step of calculation, we remind that the effective potential is
the generating functional of the one-particle irreducible Green functions at
zero external momenta. So, to have $\Pi(0)$ we can just calculate the second
derivative with respect to $\phi$ of the potential $V^{(1)}(H,T,\phi)$ in the
limit of high temperature, $ T >>\phi , T >> (eH)^{1/2}$, and then set $\phi = 0$.
This
limit can be calculated by means of the Mellin
transformation technique (see, for instance, \cite{Ska3}) and the result looks as
follows:
\begin{eqnarray} \label{22} V^{(1)}(H,\phi,T)_{\mid_{T\rightarrow \infty}} &=&
\left.[\Bigl(
\frac{C_f}{6}\phi^2_c + \frac{\alpha\pi}{2 cos^2\theta_w}\phi^2_c + \frac{g^2}
{16}\phi^2_c \Bigr) T^2  \right.]
\nonumber\\
&+& \left.[ \frac{\alpha \pi}{6} (3\lambda\phi^2_c - \delta^2(0))T^2 - \frac{
\alpha}{cos^3\theta} \phi^3 T - \frac{\alpha}{3} (\frac{3\lambda\phi^2_c -
\delta^2(0)}{2})^{3/2} T \right.]
\nonumber\\
&-& \frac{1}{2\pi} (\frac{1}{4}\phi^2_c + gH)^{3/2} T + \frac{1}{4\pi} eH T
(\frac{1}{4}\phi^2_c + eH )^{1/2}
\nonumber\\
 &+& \frac{1}{2\pi} eH T (\frac{1}{4} \phi^2_c -
eH )^{1/2}.
\end{eqnarray}
The parameter $C_f = \sum\limits_{i=1}^{3} G^2_{il} + 3\sum\limits_{i=1}^{3}
G^2_{iq}$ determines the fermion contribution of  order $\sim T^2$ having   
relevance to our problem. We also have omitted $\sim T^4$ contributions to the EP.
The terms  of the type $\sim log [T/f(\phi,H)]$ cancel the logarithmic terms in   
the temperature independent parts (\ref{11}), (\ref{18}). Considering
the high temperature limit we restrict ourselves to linear and quadratic in $T$
terms, only.

One else important expression, which also should be taken into account, is   the
linear
in $H$ term of the zero temperature EP which looks as follows:
\begin{equation} \label{lh}
V^{(1)}_{f,l}(H,\phi_c)/H_0^2 = - \frac{\alpha}{2\pi} \phi^2 \sum\limits_f K_f \mid
q_f H \mid.
\end{equation}
It significantly affects symmetry behaviour and contributes to the Debye mass in
strong
fields.

Now, differentiating these expressions twice with respect to $\phi$ and setting
 $\phi=0$, we obtain
\begin{eqnarray} \label{23} \Pi_{\phi}(0) &=& \frac{\partial^2 V^{(1)}
(\phi,H,T)}{\partial \phi^2} \mid_{\phi=0}
\nonumber\\
&=& \frac{1}{24 \beta^2}\Bigl( 6\lambda + \frac{6 e^2}{\sin^2 2\theta_w}
+ \frac{3 e^2}{\sin^2 \theta_w} \Bigr) \nonumber\\
&+& \frac{2\alpha}{\pi} \sum\limits_{f}\Bigl[ \frac{\pi^2 K_f}{3\beta^2} - \mid q_f H
\mid K_f \Bigr] 
\nonumber\\
&+& \frac{(eH)^{1/2}}{8\pi \sin^2\theta_{w}\beta} e^2 (3\sqrt{2}
\zeta(-\frac{1}{2},\frac{1}{2}).
\end{eqnarray}
The terms $\sim T^2$ give standard contributions to
temperature mass squared coming from the boson and fermion sectors.
The $H$-dependent term is negative since the difference in the brackets is
$3\sqrt{2}\zeta (-\frac{1}{2},\frac{1}{2}) - 1 \simeq - 0,39$. Formally, this
may result in the negativeness of $\Pi(0)_{\phi}$ for very strong fields
$(eH)^{1/2} > T $. But this happens in the range of parameters where asymptotic
axpansion is not applicable. Substituting expression (\ref{23}) into Eq. (\ref{21})
we  obtain (in the dimensionless variables)
\begin{equation} \label{24} {\cal V}^{\phi}_{ring} = - \frac{\alpha}{3B}
\Bigl\{(\frac{3\phi^2 - 1}{2} K  + \Pi_{\phi}(0) \Bigr\}^{3/2} + \frac{\alpha}
{3B} K (\frac{3 \phi^2 - 1}{2})^{3/2}.
\end{equation}
As one can see, the last term of this expression cancels  the  fourth term in Eq.
(\ref
{22}), which becomes imaginary at $3\phi^2 < 1$. This is the  important
 cancellation preventing  the infrared instability  at high temperature.

 Notice that  Eq. (\ref{22}) contains  other term (the last one) 
which becomes imaginary  for strong  magnetic fields or small $\phi^2$.  
It reflects the known instability  in the $W$-boson spectrum \cite{Ska1},
\cite{Niol}.

\subsection*{5.2  The Debye masses of  photons and Z-bosons }
 
To find the $H$-dependent Debye masses of photons and $Z$-bosons the
following procedure will be used. We calculate the one-loop contributions to the EP
due to the
$W$-bosons and the fermions in a magnetic field and some  ``chemical potential",
$\mu$, 
which plays the role of  an auxiliary parameter.
Then, by differentiating them twice with respect to $\mu$ and setting $\mu = 0$
the mass squared $m^2_D$ will be obtained. Let us outline that in more detail for the case 
of fermion contributions where the result is well known.

The temperature dependent part of the one-loop EP of constant magnetic field at a non-zero
 chemical potential induced by an electron-positron vacuum
polarization is \cite{Elm}:
\begin{equation} \label{25} V^{(1)}_{ferm.} =
  \frac{1}{4\pi^2} \sum\limits_{l=1}^{\infty}
(-1)^{l+1}\int\limits_{0}^{\infty} \frac{ds}{s^3} exp(\frac{-\beta^2 l^2}
{4s} - m^2s ) (eHs) coth(eHs) cosh(\beta l\mu),
\end{equation}
where $m$ is the electron mass, $e = g sin \theta_w$ is the electric charge and 
s-representation is used. Its second derivative with respect to $\mu$
taken at $\mu = 0$ can be written in the form
\begin{equation} \label{26}  \frac{\partial^2 V^{(1)}_{ferm.}}{\partial \mu^2}=
\frac{eH}{\pi^2}\beta^2 \frac{\partial}{\beta^2}\sum\limits_{l=1}^{\infty}
(-1)^{l+1}\int\limits_{0}^{\infty} \frac{ds}{s} exp(-m^2s -\beta^2 l^2/4s)
coth(eHs).
\end{equation}
Expanding $coth (eHs) $ in series and integrating over $s$ in accordance with
formula (\ref{14}) we obtain in the limit of $ T >> m, T >> (eH)^{1/2}$:
\begin{equation} \label{27} \sum\limits_{l=1}^{\infty} (-1)^{l+1} [\frac{8m}
{\beta l} K_1(m\beta l) + \frac{2}{3}\frac{(eH)^2 l\beta}{m} K_1(m\beta l)+
\cdots ].
\end{equation}
Series in $l$ can easily be calculated by means of the Mellin transformation
(see Refs. \cite{Weld}, \cite{Ska3}, \cite{SVZ2}). To have the Debye mass squared it 
is necessary to differentiate Eq. (\ref{26}) with respect to $\beta^2$ and  take into
account the relation of  the parameter $\mu$ with the zero component of the
electromagnetic potential : $\mu \rightarrow  ieA_0$ \cite{SVZ1}. In this way we obtain finally,
\begin{equation} \label{28} m^2_{D} = g^2 sin^2\theta_w ( \frac{T^2}{3} -
\frac{1}{2\pi^2} m^2 + O((m\beta)^2, (eH\beta^2) ) ).
\end{equation}
This coinsides with the known result calculated from the photon polarization operator
 \cite{VZM}. As  one can see, the dependence on $ H$ appears in the order
$\sim T^{-2}$. To find the total fermion contribution to $m^2_D$ one has to sum up 
the expression (\ref{28}) over all fermions and substitute their electric charges.

To calculate $m^2_D$ for $Z$-bosons it is sufficient to account of the
fermion coupling to $Z$-field. It can be done by substituting $ \mu \rightarrow
i(g/2 cos \theta_w + g sin^2 \theta_w ). $ The result differs from Eq.
(\ref{28}) by the coefficient at the brackets in the right-hand side which
has to be replaced, $g^2 sin^2 \theta_w \rightarrow g^2 (\frac{1}{4 cos^2 \theta_w} 
+ tang^2 \theta_w) $.
 One also has to add the terms coming due to the neutral currents and the part of 
fermion-Z-boson interaction which is not reproduced by the above substitution:
\begin{equation} m^{2'}_{D} = \frac{g^2}{8 cos^2 \theta_w} (1 + 4 sin^4
\theta_w) T^2 .
\end{equation}
\nonumber\\

Now, let us apply the above procedure to obtain the $W$-boson contribution.
The corresponding EP (temperature dependent part) calculated at non-zero $T, \mu
$ looks as follows,
\begin{equation} \label{29} V^{(1)}_w = - \frac{eH}{8\pi^2}\sum\limits_{l=1}^{
\infty}\int\limits_{0}^{\infty} \frac{ds}{s^2}exp(-m^2 s -l^2\beta^2/4s)[\frac{
3}{sinh(eHs)}+ 4 sinh(eHs)] cosh(\beta l\mu).
\end{equation}
All the notations are obvious. The first term in the squared brackets gives
the triple contribution of the charged scalar field and the second one is due
to the interaction with a $W$-boson magnetic moment. Again, after
differentiation twice with respect to $\mu$ and setting $\mu = 0$ it can be
written as
\begin{equation} \label{30} \frac{\partial^2 V^{(1)}_w}{\partial \mu^2}_{\mid \mu =0} =
\frac{eH}{2\pi^2}\beta^2\frac{\partial}{\partial\beta^2} \sum\limits_{l=1}^
{\infty}\int\limits_{0}^{\infty}\frac{ds}{
s}exp(- \frac{m^2s}{eH} - \frac{l^2\beta^2eH}{4s})[\frac{3}{sinh(s)} + 4
sinh(s)].
\end{equation}
Expanding $sinh^{-1}s$ in series over Bernoulli's polynomials,
\begin{equation} \frac{1}{sinh s} = \frac{e^{-s}}{s} \sum\limits_{k=0}^{\infty}
\frac{B_k}{k!}(-2s)^k,
\end{equation}
\nonumber\\
and carrying out all the calculations described above, we obtain for the $W$-
boson contribution to $m^2_D$ of the electromagnetic field
\begin{eqnarray} \label{31} m^2_D &=& 3 g^2 sin^2 \theta_w [ \frac{1}{3} T^2 -
\frac{1}{2\pi} T(m^2 + g sin\theta_w H)^{1/2} - \frac{1}{8\pi^2}(g sin\theta_w H)
\nonumber\\
&+& O(m^2/T^2, (g sin\theta_w H /T^2)^2)].
\end{eqnarray}
Hence it follows that  spin does not affect the Debye mass in leading order.
        Other interesting feature is that the next-to-leading terms are negative. 

The contribution of the $W$-boson sector to the $Z$-boson mass $m^2_D$ is given by 
expression (\ref{31}) with the replacement $g^2 sin^2 \theta_w \rightarrow g^2 cos^2 \theta_w$.
Summing up the expressions (\ref{28}) and (\ref{31}) and substituting them in
Eq. (\ref{21}), we obtain the photon part $V^{\gamma}_{ring},$ where it is
necessary to express masses in terms of the vacuum value of the scalar
condensate $\phi_{c}$.
 In the same way the daisy diagrams of $Z$-bosons $V^{z}_{
ring}$ can be calculated. The only difference is the mass term of $Z$-field  and the
additional term in the Debye mass due to  the neutral current
$\sim \bar{\nu}\gamma_{\mu}\nu Z_{\mu}$. These three fields - $\phi, \gamma,
 Z$,- which become massless in the
restored phase, contribute into $V_{ring}(H, T)$ in the presence of the magnetic
field. At zero field, there are also terms due to the $W$-boson loops
in Figs. 3, 4. But when $H \not = 0$ the charged  particles acquire masses $\sim
eH$
and these daisies can be neglected.

\subsection*{5. 3  Daisy diagrams of the tachyonic mode }

A separate consideration should be spared to the tachyonic (unstable)  mode in the
$W$-boson spectrum: $p^2_0 = p^2_3 + M^2_w - eH$.  First  of all, we  notice that
this mode is excited due to a spin interaction and it does not influence the
$G_{00}(k)$ component of the $W$-boson propagator.  Secondly,  in the fields
$eH \sim M^2_w$ the  mode becomes the long range state. Therefore,  it should be
included in  $V_{ring}(H,T)$ side by side  with  other considered neutral fields.  But in
this case
it is impossible  to take advantage of formula (\ref{21}) and we return to the
initial
EP containing the generalized propagators.

For our purpose  it will be convenient to make use of the  generalized EP
written as the sum over the modes in the external magnetic field \cite{SVZ1},
\cite{SVZ2}:
\begin{equation} \label{TDVZ}  V^{(1)}_{gen} =  \frac{eH}{2\pi \beta} \sum
\limits_{l= -\infty}^{+ \infty} \int\limits_{- \infty}^{+ \infty} \frac{dp_3}{2\pi}
\sum\limits_{n = 0, \sigma = 0,\pm 1}^{\infty} log [\beta^2(\omega^2_l + \epsilon   
^2_{n,\sigma,p_3} + \Pi(T,H) )] ,
\end{equation}
where $\omega_l = \frac{2\pi l}{\beta}$ ,  $\epsilon^2_n = p_3^2 + M^2_w + (2n + 1 -
2\sigma) eH $ and $\Pi(H,T)$ is the radiation mass squared of  $W$-bosons  in  magnetic
field at finite temperature.
Denoting   as $D^{- 1}_0(p_3,H.T)$ the sum $ \omega^2_l + \epsilon^2$, one can
rewrite eq. (\ref{TDVZ}) as follows:

\begin{eqnarray} \label{TDVZ1}
V^{(1)}_{gen} &= & \frac{eH}{2\pi\beta}
\sum\limits_{l=-\infty}^{+\infty}\int\limits_{-\infty}^{+\infty} \frac{d p_3}{2\pi}
\sum\limits_{n,\sigma} log[\beta^2 D^{- 1}_0(p_3,H,T)]
\nonumber\\
&+&  \frac{eH}{2\pi\beta} \sum\limits_{l = - \infty}^{+ \infty}  \int\limits_{-
\infty}^{+\infty} \frac{d p_3}{2\pi} \{ log[ 1 + ( \omega^2_l + p^2_3 + M^2_w - eH)^{-
1} \Pi(H,T) ] \nonumber\\
&+& \sum\limits_{n \not = 0, \sigma \not = +1 } log[ 1 +  D_0 ( \epsilon^2_n , H,T) 
\Pi(H,T) ] \}.
\end{eqnarray}
Here,  the first term  is just the one-loop contribution of $W$-bosons,  the second one
gives the  sum of daisy diagrams  of the unstable mode (as it can easily be checked 
by
expanding the logarithm in a series). The last term describes the  sum of the short
range modes in the magnetic field and should be omitted.

Thus, to find  $V^{unstable}_{ring}$ one has to calculate the second term in Eq.
(\ref{TDVZ1}). In the high temperature limit we obtain:
\begin{equation} \label {Vunst} V^{unstable}_{ring} = \frac{eH}{2\pi\beta} \{ (
M^2_w - eH + \Pi (H, T) )^{1/2} - ( M^2_w - eH )^{1/2} \}.
\end{equation}
By summing up  the one-loop EP and   all the  terms $ V_{ring}$, we arrive at the total
 consistent in leading order EP.

Let us mention the most important features of the above expression. It is seen that the
last
term in Eq. (\ref{Vunst}) exactly cancels the "dangerous" term in Eq. (\ref{22}). So,
 the EP is real and no instabilities appear at sufficently high temperatures when $\Pi  
( H,T) > M^2_w - eH $.  To make a quantitative estimate of the range of validity of the 
total EP
it is necessary to calculate the $W$-boson mass operator in a magnetic field at finite
temperature and hence to find $ \Pi (H,T)$. This is a separate and enough cogent
problem which is considered in  detail in Ref. \cite{SS}. Here, we only adduce
the result
of $\Pi (H,T)$ calculations:  
\begin{eqnarray} \label{munst1}
\Pi_{unstable}(H, T) &=&  <n=0,\sigma=1\mid \Pi^{charged}_{\mu \nu} \mid n=0,
 \sigma = 1>\nonumber\\
      &=&  \alpha [12, 33 (eH)^{1/2} T  +  i 4 (eH)^{1/2}T] ,
\end{eqnarray} 
where  the average value of the mass operator in the ground state of the $W$-boson spectrum
$ \mid n =0, \sigma = + 1> $  was calculated. This expression has been obtained in
the limit $eH/ T^2 << 1, B = M_w/T << 1$.  Side by
side with 
the real part responsible for the radiation mass squared 
the expression (\ref{munst1}) contains the imaginary one describing the decay of the
state. The latter term is small as the former one is compared and of  order of the  usual
damping constants at high temperature. So, $Im \Pi (H,T)$ can be ignored in our
problem. The radiation mass  squared is positive. It acts  to stabilize the spectrum.
At $H = 0$ no screening is produced in one-loop order, as it should be at finite
temperatures for transversal modes \cite{Kal}. Thus, we come to conclusion that at
sufficiently
high temperature the effective $W$-boson mass squared $M^2_{w_{eff.}} = M^2_w
 - eH + \Pi (H,T)$ is positive and no conditions for $W$-boson condensation
discussed in Refs. \cite{Sk1}, \cite{AO1} are realized.
\footnote{  Express  (\ref{munst1}) disagrees with the corresponding  one of Ref.
\cite{Elmp} where the average value of the gluon polarization operator in an abelian
chromomagnetic field was calculated in weak field approximation and $\Pi(H,T)$ has
been found to be zero. 
Most probably,  the discrepancy is the concequence of the calculation procedure adopted by
these authors when the gluon polarization operator was calculated at zero external
field and  then its average value has been calculated in the state $\mid n = 0, \sigma
 = + 1> $ . Our expression is the high temperature limit of the mass operator which
 takes into account the external field exactly.}

\subsection * {6. RESTORED PHASE IN THE EXTERNAL MAGNETIC FIELDS}

Having calculated the EP at $\phi \not  = 0$ we are able to  determine the kind of
the EW
phase transition
 for different $m_H, h$. That will be done in the next sections. Here, we are
 going to describe in more detail the restored phase with the hypermagnetic
 and magnetic fields. Let us consider first the former case. 

 To describe more precise the restored phase one has to calculate radiation
 corrections to
 the external field $H_Y$ at high temperature. Before doing that let us remind
 that at $\phi = 0$ this field is completely unscreened whereas the nonabelian
 constituents of the electromagnetic and $Z$-fields are screened at scales $l \geq
  (g^2 T)^{-1}$ by the magnetic mass. Remind also that we are investigating the
separate influence
of the external fields. This
   means that in the covariant derivative describing interaction with
 the external field $H_Y$  in the restored phase one should maintain
 the $U(1)_Y$ term only:
  $D_{\mu} = \partial_{\mu} + \frac{1}{2} g^{ } B_{\mu}^{ext}$. We set the
   potential as before, $B_{\mu}^{ext} = ( 0, 0, H_Y x^{1}, 0 )$.

In the restored phase $W$-bosons do not interact with $H_Y $.  The field dependent part
of the EP $V ( \phi = 0, H_Y, T )$ is non-zero due to the contributions of fermions and
scalars. However, the fermion part depends logarithmically on temperature ($ \sim
\frac{(g'/2)^{2}}{4 \pi} H^2_Y log T/ T_0 $) and can be neglected as compared to the tree
level term $\frac{1}{2} H_Y^2$. This is not the case for  the scalar field whose  
contribution to  the one-loop EP is
\begin{eqnarray} \label{Yscal}
V^{(1)}_{sc}(H_Y, T ) =  &-&\frac{(g'/2)^2 H_Y^2}{24 \pi^2} ln (T/T_0)\nonumber\\
 &+&
\frac{((g'/2) H_Y)^{3/2} T}{ 6 \pi} + O (1/T).
\end{eqnarray}
The term logarithmically dependent on $T$ can again be neglected but the linear in $T$
part should be retained. Since ``hyperphotons'' are massless in the restored phase we
also
include the contribution of the corresponding daisy diagrams:
\begin{equation}  \label{Y1ring} V_{restored}^{ring}(H_Y, T ) = - \frac{T}{12 \pi}
[\frac{2}{3} (g'/2)^{2} T^2 +
m^2_{D_f} - \frac{((g'/2) H_Y)^{1/2} T}{2 \pi} - \frac{1}{8 \pi^2} (g'/2)H_Y ]^{3/2},
\end{equation}
where $m^2_{D_{f}}= \frac{1}{24} g'^2 T^2 \sum\limits_{f(R,L)}Y^2_f$ is the sum over
the fermion
contributions to the Debye mass of the ``hyperphotons'', $Y_f$ are the hypercharges  of $R-$
and $L-$ leptons and quarks.
  Both these expressions have been calculated in a way described in previous sections.

For convenience of numerical investigations we  express Eqs. (\ref{Yscal}) and
(\ref{Y1ring}) in terms of the dimensionless variables $h,~ B $: $V(H_Y, T )_{restored}   
= (H_0)^2 v_{restored} (h, B )$,
\begin{eqnarray} \label{EPr} v_{restored} (h, B ) &=&\frac{1}{2} \frac{h^2}{
cos^2 \theta } +  \frac{\alpha}{3\sqrt{2} cos^3 \theta} \frac{h^{3/2}}{B}
\nonumber \\
&-& \frac{1}{3}\frac{\alpha}{B} [\frac{7}{6} \frac{4 \pi \alpha}{ cos^2 \theta B^2}  -
\frac{h^{1/2}}{2\sqrt{2}\pi B cos\theta} - \frac{h}{16\pi^2 cos^2\theta}]^{3/2},
\end{eqnarray}
where $\alpha = e^2/ 4 \pi$ and $h_Y = h/cos\theta$.

Now, let us turn to the magnetic field case.  In accordance with our approach   we  set
$H_Y= 0$. It will be important for
        what follows to remember recent results on  obeservation of the gluon magnetic
        mass in lattice simulations that was found to be of order $m_{mag} \sim g^2
        T$ (as it has been expected from nonperturbative calculatios in quantum field
        theory \cite{Kal}, \cite{Buch}). The mass screens  the nonabelian component of
the magnetic
        field at distancies $l > l_m \sim (g^2 T)^{-1}$ but inside the space domain $l
        < l_m$  it may  exist and affect all the processes at high
        temperatures, as it was discussed in sect. 2.1. This fact has not been taken
into consideration in a number of
        investigations of the EW phase transition. In particular, in Ref. \cite{Elmp}
(as  
in Ref. \cite{Enol}) the field strength generated at finite temperature was
        erroneously estimated as coinsiding with that at zero temperature.

The magnetic field in the restored phase may be homogeneous or not dependently on  the
      stability of the charged boson spectrum in the field  calculated with radiation
      corrections included. 
That can be checked in accordance with formulae (\ref{Vunst}) and (\ref{munst1})
at $M_w =
0$. If the effective mass squared $M^2_{eff.}(H, T)$ is positive, the perturbative vacuum is
stable and the external field is homogeneous otherwise it is unstable and a lattice
structure has to be generated due to the evolution of the instability. We shall see
below, for the first order phase transition (which is the
main topic in the present paper) $M^2_{eff.}(H, T,
\phi_c)$ is positive in the minimum of the EP when the symmetry restoration happens.
Therefore, we will not investigate in detail the structure of the restored phase for
different $H$ restricting ourselves by the case of the constant field, as in the broken
phase.

To make a link between studies of symmetry behaviour in the external hypermagnetic field
  and the previous results for the  usual magnetic field \cite{Rez},
\cite{SVZ1}
       we notice that  in the broken phase  $H_Y$ and $H$ are connected by the
relation
       $H = H_{Y} \cos \theta$. So, all
       investigations, dealing with symmetry behaviour in a magnetic field at high
       temperature, are relevant in the case of $H_Y$ in the respect of the form of the EP
       curve at different $T, H_Y$. The hypercharge field influences the scalar field
       condensate at tree level and acts to restore symmetry. That was the reason why it
       has been taken into account in the lower order in Refs. \cite{Shap1}, \cite{Elm1}.
But, as it
       will be shown below,  for strong fields and heavy $m_H$ the form of the EP
       curve in the broken phase is very sensitive to the change of the parameters.
       Moreover, it is strongly depended on the correlation correction contributions.
       So, to have an adequate picture of the  EW phase transition
       symmetry behaviour with the daisy diagrams included has to be investigated.

\subsection*{7. HIGH TEMPERATURE EXPANSION OF THE EFFECTIVE POTENTIAL}

The most of investigations dealing with symmetry behaviour at high temperature used 
the limiting form of the EP at $T \rightarrow \infty$. It will be of interest 
to compare the results obtained in  two cases - for the 
complete EP and for the asymptotic one.

First, let us adduce the high temperature limit of the sum of terms describing 
the contributions of the scalar field at zero and finite temperature, the 
$H$-independent contribution of $W$-bosons, $Z$-bosons as well as the contributions of
 the second terms of corresponding daisy diagrams, Eq. (\ref{24}):
\begin{eqnarray}\label{V*1}
\omega &=& \frac{3\alpha}{8\pi} \frac{1}{cos^4\theta}( \phi^2 + \phi^4 
(log(\frac{4\pi cos \theta}{B})^2 - C )) \\
\nonumber
&+& \frac{3 \alpha}{4\pi} (\frac{1}{2} \phi^4
log \phi^2 - \frac{3}{4} \phi^4 + \phi^2 )  \nonumber\\
&+& \frac{\alpha K^2}{32\pi}[ \frac{1}{2} (3 \phi^2 - 1)^2 log\frac{(4\pi)^2}{B^2 K} 
+ 6 \phi^2 -C (3 \phi^2 - 1)^2 ] \\
\nonumber
&+& \frac{\alpha}{\pi}[ \frac{\pi^2 K}{12 B^2}(3 \phi^2 - 1) 
+ \frac{\pi^2 \phi^2}{2 cos^2\theta B^2} - \frac{2\pi\phi^3}{3\cos^3\theta B}],
\end{eqnarray}
$C$= 0.5772 is Euler's constant.   The first terms of daisies dependent on the
Debye masses should be added separately.

The high temperature asymptotic of the fermion sector $\omega_f$ looks as folows:
\begin{eqnarray} \label{V*f}
\omega^{*}_f &=& -\frac{\alpha}{\pi} \sum\limits_f [ - \frac{\pi^2}{3} \frac{K_f\phi^2}
{B^2} + \frac{1}{6}q_f^2 h^2 (log(\frac{\pi}{4\pi \alpha + K_f \phi^2 B^2}) - 2 C )\\
\nonumber 
&+& \frac{K^2_f \phi^4}{4} (log(\frac{\pi}{4\pi\alpha + K_f\phi^2 B^2}) - 2 C 
+ \frac{3}{4})].
\end{eqnarray}
 The term $4\pi \alpha $ in Eq. (\ref{V*f}) is introduced in order to correct 
the infrared infinity in the field. It effectively accounts of the fermion
 temperature mass. Just due to this infrared singularity   the light fermions dominate in
strong fields.

Now, let us wright down the high temperature limits of the $W$-boson sector
(expressions 
$\omega_0,~  \omega_1 - \omega_2)$ Eq. (\ref{14a}). Instead Re $\omega^{(1)}_w$ one has
to
substitute
 the sum $\omega_{spin} + \omega_0 = \omega^{'}_w$:
\begin{eqnarray} \label{V*w}
\omega_{spin} &=&- \frac{\alpha h}{2\pi}[ (\phi^2 + h)log(\phi^2 + h) - (\phi^2 - h)
log\mid \phi^2 - h \mid  \\ \nonumber
 &-&2 h + 8 C h - 2 h log(\frac{4\pi}{\mu^2 B^2}) + \frac{4\pi}{B}( \phi^2 + h )^{1/2} ],
\end{eqnarray}
\begin{eqnarray} \label{V*0}
\omega_0 &=& -\frac{3\alpha}{\pi} [ \frac{h^2}{12} (log\frac{4\pi}{(h 
+ \phi^2)^{1/2}B} - C ) - \frac{\phi^4}{4}(log(\frac{4\pi}{(h 
+ \phi^2)^{1/2}B}) - C )\\\nonumber
&-& \frac{\pi^2 \phi^2}{3 B^2} + \frac{2\pi}{3B} (\phi^2 + h)^{3/2} 
- \frac{\pi h}{B} (h + \phi^2)^{1/2} + \frac{1}{16}h^2 - \frac{1}{8}\phi^2 h - \frac{3}{16} \phi^4 ].
\end{eqnarray}
Here, the term due to the unstable mode is omitted since it is cancelled by the second
 term of daisy diagrams  generating by the unstable mode. As it is seen, a lot of terms
 from expresion $V_{z,\phi}$ at zero temperature are cancelled in the total. 
The term with $log (B\mu)$ is $\phi$-independent and therefore inessential when
 symmetry behaviour is investigated, $\mu$ marks the normalization point.

Now, for completeness, let us write down explicitely the contributions of the 
$\gamma$- and $Z$- daisy diagrams:
\begin{equation} \label{r3}
V_{ring}^{\gamma} = - \frac{\alpha}{3 B}[\Pi_{\gamma}(h,B)]^{3/2},
\end{equation}
where
\begin{eqnarray} \label{r2}
\Pi_{\gamma}(h,B) &=&  \sum\limits_f ( \frac{e^2 q^2_f}{3B^2} 
- \frac{e^2 q^2_f K_f \phi^2}{2\pi^2} )\\
\nonumber
&+& e^2 (\frac{1}{B^2} - \frac{3}{2 \pi B} (\phi^2 + h )^{1/2} 
- \frac{3}{8\pi^2} h ).
\end{eqnarray}

As the contribution of the $Z$ boson daisy diagram we get (taking into account 
the subsitution  $e^2 \rightarrow g^2 (\frac{1}{4 cos^2 \theta} + tan^2 \theta)
 $ and addind other necessary terms):
\begin{equation} \label{r4}
V_{ring}^{z} = - \frac{\alpha}{3 B}[\Pi_{z}(h,B) + \frac{\phi^2}
{cos^2 \theta}]^{3/2} + \frac{\alpha}{3B} \frac{\phi^3}{cos^3\theta},
\end{equation}
where
\begin{eqnarray} \label{pz}
\Pi_{z}(h,B) &=&  \sum\limits_f( g^2 (\frac{1}{4 cos^2 \theta} + tan^2 \theta)
 ( \frac{ q^2_f}{3B^2} - \frac{ q^2_f K_f \phi^2}{2\pi^2} ))\\ \nonumber
&+& \frac{g^2}{8 cos^2 \theta} ( \frac{1}{6B^2} - \frac{\sqrt{K} \phi}{2\pi B})\\
\nonumber
&+& g^2 (\frac{1}{4 cos^2 \theta} + tan^2 \theta ) (\frac{1}{B^2} - \frac{3}{2 \pi
B} (\phi^2 + h )^{1/2} - \frac{3}{8\pi^2} h )\\ \nonumber 
&+& \frac{g^2}{8 B^2 cos^2 \theta}(1 + 4 sin^4 \theta).
\end{eqnarray}
 The term, containing $\sqrt{K}$ comes from the one-loop diagram of the Higgs
field.  
Remind, for the case of the asymptotic EP the last term of Eq. (\ref{r4})
 has already been included in Eq. (\ref{V*1}).

\subsection*{8. SYMMETRY BEHAVIOUR IN STRONG HYPERMAGNETIC FIELD}

Let us investigate the EW phase transition in the hypermagnetic field for different 
values of $m_H$. It can be done by considering the Gibbs free energy in the broken,
 $G_{broken}(H^{ext}, \phi_c, T),$ and the restored, $G_{restored}(H_Y^{ext}, T)$,
 phases \cite{Shap1}, \cite{Elm1}:
\begin{eqnarray} \label{Gib}
G_{broken} &=& V(\phi, H,T) - \vec {H}\vec{H}^{ext}_Y  cos \theta,\\
\nonumber
G_{restored} &=&  V(0, H_Y,T) - \vec{H_Y}\vec{H}_Y^{ext}. 
\end{eqnarray}
By minimization of these equations the fields $H$ and $H_Y$ generated in the vacuum 
have to be expressed throught $H_Y^{ext}$.  The first order phase transition can be 
determined within two equations:
\begin{equation} \label{cond}
G_{restored}(H_Y^{ext}, T, 0) = G_{broken}( H^{ext}, T, \phi(H^{ext})_c ), 
\end{equation}
describing the advantage of the broken phase  creation, 
where $\phi(H)_c$ is a scalar field vacuum expectation value at given $H, T,$ which has 
to be found as the minimum position of the total EP,
\begin{equation} \label{min}
\frac{\partial V(H, T, \phi_c)^{total}}{\partial \phi_c} = 0.
\end{equation}
Hence the critical field strength can be calculated. In this expression and below we
 use for brevity $H$ instead $H^{ext}$.

Having obtained the EP in the restored phase, the one-loop EP described by 
formulae (\ref{12}), (\ref{14a}),
(\ref{16})-(\ref{20}) and the daisy diagram contributions  $V_{ring}$ we are
going to investigate symmetry behaviour. We shall present the results  in two 
stages. First, we shall consider the total EP as the function of $\phi^2$ at various 
fixed $H$, $T$, $K$ and determine the form of the EP curves in the broken phase. 
In this way it will be possible to select the range of the parameters when the first
 order phase transition is realized. After that the 
temperature  $T_c$ at given field strength $H_Y$ will be found.

As usually \cite{Sk2}, to investigate symmetry behaviour we consider the 
difference ${\cal V^{'}} = Re [{\cal V}(h,\phi, K, B) - {\cal V}(h, \phi= 0, B )]$  
which  gives information about the symmetry restoration. We will present the results
for
 two cases: 1) for the total EP in the broken phase; 2) for the high temperature limits
 of it. Then we will make a  comparison.

In what follows, it will be also convenient to express the conditions of the phase 
transition in terms of the dimensionless variables $h, B, \phi,$ taking into account
 the relation $h_Y = h/cos \theta$. Then,  the Gibbs free energy 
\begin{equation} \label{Gbrok}
G_{broken}(h^{ext}, \phi, B) = \frac{h^2}{2} + v^{'}(h, \phi, B) - h h^{ext},
 \end{equation}
has to be expressed in terms of $h^{ext}$ through the equation
\begin{equation} \label{cond}
h^{ext} = h + \frac{\partial v^{'}(h, \phi, B)}{\partial h},
\end{equation}
where $v^{'}$ describes the one-loop and daisy diagram contributions to the EP. 
The phase transiton happens when the condition
\begin{equation} \label{Tc}
\frac{h^2}{2}tan^2\theta = v^{'}_{restored}(h, B_c) - v^{'}_{broken}(h, \phi_c, B_c)
\end{equation}
 holds.
The function $v_{restored}^{'}$ is given by Eq. (\ref{EPr}). We also have substituted
 the field $h^{ext}$ by $h$.

The results on the phase transition determined
by numeric investigation of the total EP are summarized in Table 1.
~ \\ ~  

\begin{tabular}{c c c c c c c} \hline
~ \\ ~
~h&$K$&$T_c(GeV)$&$\phi_c(h,T_c)$&$\phi^2_c(h,T_c)$&$R$&$M_w(h,T_c)$ \\
[5pt] \hline
~0.01&0.85&106.47& 0.301662 & 0.091 & 0.69699 & 0.327235 \\
~0.01&1.25&122.21& 0.181659 & 0.033 & 0.36567 & 0.230086 \\
~0.01& 2  &145.56& 0.094868 & 0.009 & 0.16033 & 0.186168 \\ \hline

~0.1 &0.85&108.58& 0.275681 & 0.076 & 0.62459 & 0.245186 \\
~0.1 &1.25&123.54& 0.130384 & 0.017 & 0.25963 & 0.112721 \\
~0.1 & 2  &148.39& 0.031623 & 0.001 & 0.05242 & 0.126315 \\ \hline   

~0.5 &0.85&108.89& 0.248998 & 0.062 & 0.56253 & 0.49938 i\\
~0.5 & 1.25&second&order & phase &transition                         \\
~0.5 & 2   &second&order & phase &transition          \\ \hline
~ \\ ~
Table 1.
\end{tabular}

In the first column we show the hypermagnetic field strength in the
broken phase (in dimensionless units). In the second and third ones the
mass parameter $K = m^2_H/ M^2_w$ and the critical temperature of the first order
phase transition are adduced. Next two columns give the local minimum position
$\phi_c(H, T_c)$  and its squared value at the
transition temperature. The last two columns fix the ratio $R =
246 GeV \phi_c(h, T_c)/ T_c $, determining the advantage of
baryogenesis, and the $W$-boson effective mass calculated in the local
minimum of the EP at the corresponding field strength and the transition
 temperature. The parameter $M_w(h, T_c) = [(\frac{g}{2} \phi_c(h, T_c))^2 - eH +
\Pi(H, T_c, \phi_c)]^{1/2} /M_w$ is the dimensionless W-boson mass with the radiation
corrections included.  

As it is seen, the increase in $h$ makes the phase
transition weaker (not stronger as it was expected in Refs. \cite{Shap1}
, \cite{Elm1} by analogy to superconductivity in the external magnetic
field). The ratio R is less then unit for  all the field strengths,
wherease the baryogenesis condition is $R > 1.2 - 1.5$ \cite{Shap2}.
Thus, we come to the conclusion that external hypermagnetic field does
not make the EW phase
transition strong enough to produce baryogenesis.
Moreover, for strong fields the phase transition is of second order
 for all the values of $K$ considered.

 Let us continue the analysis of data in the Table 1. For the field
strengths $H > 0.1 - 0.5 H_0 (H_0 = M^2_w/e)$ the phase
transition is of  second or weak first-order. The W-boson effective mass
squared
(in dimensionless units) $M^2_w(\phi_c, h, B_c) = \phi^2_c
(h, B_c) - h + \Pi(h, B_c)$ is positive for $h = 0.01$ and $h = 0.1$.   
Therefore, the local minimum is the stable state at the first order phase
transition. For stronger fields, when the second order phase transition
 happens, the effective $W$-boson mass becomes imaginary. This reflects
the known instability in the external magnetic field which exhibits
itself even when the radiation mass of the tachyonic mode is included.
 But it does not matter for the problem of searching for the strong first
order phase transition in the external field investigated in the present
paper. The instability has to result in the condensation of $W$- and
$Z$-boson fields at high temperature. However, this does not change the type of the
phase transition. 

In Refs. \cite{Shap1}, \cite{Shap4}, \cite{Elm1}, \cite{Lain}
it has been concluded that the strong hypermagnetic field increases the
strength of the first order phase transition and in this case
baryogenesis survives in the SM. Our results are in obvious
contradiction with this conslusion. To explane the origin
of the discrepancies let us first consider Refs. \cite{Shap1}, \cite{Elm1}
 where a perturbative method of computations has been applied. These
authors, in studying of the EW phase transition, have allowed for the
influence of the external field at tree level, only. That corresponds to the
usual case of superconductors in the external magnetic field, and, as a
consequence, they observed the strong first order phase transition.
 In fact, the type of the phase transition was just assumed, since no
investigations of the EP curve with all the particles included for different $H_Y, T$ have
been
carried out.
In the former paper there was the
qualitative estimate of the field effect, whereas in the latter
one the quantitative analysis in one-loop approximation for the temperature
dependent part of the EP has been done. Actually,  in both these
investigations the influence of the external field was reduced to
the consideration of the
condition (\ref{Tc}) fixing the transition temperature in the
hypermagnetic field. The role of fermions and $W$-bosons in the field
 was not investigated at all. However, as we have seen, the
fermions (heavy and light) are of paramount importance in the phase
transition dynamics. Just due to them the  EW phase transition becomes
of  second order in strong fields (for the values of $K$ when it is of
first order in weak fields).

In Refs. \cite{Shap4}, \cite{Lain} the phase transition was investigated
by the method combining the perturbation theory and  the lattice   
simulations.  As the first step in this approach the static modes only are 
 maintained in the high temperature Lagrangian. The fermions are nonstatic
 modes and decoupled. The only fermion remainder  is the t-quark mass entering the
effective universal theory \cite{Shap2}, \cite{Shap4}, \cite{Lain}. 
 So, no  fermion features in the external fields and
hence no information about the form of the EP curve could be derived in this way.
In our analysis, it has been   
observed that not only heavy but also light fermions
are important in strong external fields, as one, in particular, can see from the 
term  $H^2 logT/m_f$ of the one-loop EP.
At high temperature it influences symmetry
behaviour considerably. Actually, for various field strengths  the fermoins
with different masses are dominant and we have allowed for all of them. Besides, 
we have taken into consideration all the daisy correlation corrections in the
external field that also affects symmetry behaviour.

We would like to stress that our  perturbative results for
the values of $K \sim 0.8 - 0.9$ are reliable. They are in agreement with   
nonperturbative analysis at zero field. The external field is taken into consideration 
exactly. For these mass $m_H \sim 75 - 80 $ GeV we observed 
 the change of the first order phase transition into the second order one
 with increase in the field strength.
The same behaviour takes also place for $K > 1$ when the perturbative analysis may
be not trusty. But, as we have determined, the picture of the symmetry behaviour is
 only quantitatively changed for heavy scalar particles: the first order phase
transition in weak fields becomes the second order one for strong
 fields.  These circumstances convince us that the assumption of Ref.
 \cite{Shap1} that the hypermagnetic field makes the weak first-order phase transition strong
enough is not proved by the detailed   calculations.

 In Table 2 we adduce the results for the phase transition obtained within the
asymptotic EP. 
~ \\ ~

\begin{tabular}{c c c c c c c} \hline
~ \\ ~
~ h &K&$T_c(GeV)$&$\phi^2_c(h, T_c)$&$\phi_c(h, T_c)$&$R$& $M_w(h, T_c)$ \\
[5pt] \hline
~ 0.01 & 0.85 &104.678&0.097& 0.3114 &0.7319 & 0.3358 \\
~ 0.01 & 1.25 &119.569&0.098& 0.3146 &0.6473 & 0.3441 \\
~ 0.01 & 2    &142.01 &0.102& 0.3194 &0.5532 & 0.3563 \\  \hline

~ 0.05 & 0.85 &105.449&0.085& 0.2915 &0.6801 & 0.3050 \\
~ 0.05 & 1.25 &120.241&0.094& 0.3066 &0.6765 & 0.3319 \\
~ 0.05 & 2    &142.559&0.102& 0.3194 &0.5511 & 0.3611 \\ \hline

~ 0.1  & 0.85 &second &order& phase &transition   \\
~ 0.1  & 1.25 &121.616&0.062&0.2490 &0.5037  & 0.2809  \\
~ 0.1  & 2    &143.539&0.088&0.2966 &0.5084  & 0.3420   \\ \hline
~ \\ ~
 
Table 2.
\end{tabular}

These data, as previous ones, show that the ratio $R$ is less then unit and the
baryogenesis condition does not hold. 

Now, let us compare the results in Tables one and two. As one can see,
even for $h = 0.01$ they differ
considerably for all the parameters exept the critical temperatures. The first order
phase transition for $K = 0.85$ determined by both of the potentials possesses the same
characteristics. But this is not the case for $K = 1.25 , 2 $ when the phase
transition described by the exact EP is weaker first order. For the field strength $h
= 0.1$ the exact EP predicts the weak first-order phase transition for all $K$
whereas the asymptotic one fixes the second order phase transition for $K = 0.85$.
For $K = 1.25, 2$ the jump of the order parameter determined by the asymptotic EP is
 twice larger then for the exact EP case. If the fields are stronger then 0.2 - 05
$\cdot 10^{24}$ G both EP predict the second order phase transition.
Thus, we conclude that asymptotic EP predicts conversion of the first order
phase transition to the second order one for weaker fields whereas the first order
phase trasnsition described by it is stronger as copmare to the transition derived
within the exact EP.

\subsection*{9. SYMMETRY BEHAVIOUR IN  STRONG MAGNETIC FIELD}

As we have described qualitatively in sect. 2, the case of external magnetic
 field somehow differs from the hypermagnetic one.   Nevertheless, a nubmer
 of previous results dealing mainly with the form of the EP curve in the broken
 phase are relevant, since in this phase the unscreened constituent of the external
hypermagnetic field  coinsides with the magnetic field.  The condition fixing the
transition temperature is 
\begin{equation} \label{phtrH}
V_{restored}(H, T_c, 0) = V_{broken}(H, T_c, \phi_c(H, T_c)).
\end{equation}
The transition happens when the depth of the minima located at the begining, $ 
\phi_c
= 0,$ and at $\phi_c \neq 0$ is the same. 

Below, as before,  we present the results of numeric investigations of the
phase transition obtained
 for the exact EP and  for the high temperature limit of it. 
 In Table 3 we show the characteristics of
 the first order phase transition determined within the exact EP.

   ~\\~
\begin{tabular}{c c c c c c c} \hline
~\\~
~ h&$K$&$T_c(GeV)$&$\phi^2_c(h,T_c)$ &$\phi_c(h,T_c)$& $R$&$M_w(h,T_c)$ \\[5pt]
 \hline
~0.01&0.85&105.18&0.107&0.3271&0.76506& 0.3504     \\
~0.01&1.25&120.77&0.045&0.2121&0.4321 &0.2541    \\
~0.01&2   &143.96&0.016&0.1265&0.2162 & 0.2031   \\ \hline

~0.1&0.85 &106.35&0.098&0.3130&0.7241 &0.2835    \\
~0.1&1.25 &121.92&0.021&0.1449&0.292  &0.1243  \\
~0.1& 2   &146.14&0.003&0.0545&0.0917 &0.1272   \\ \hline

~0.5&0.85 &108.23&0.092&0.3033&0.6894 &0.4696 i   \\
~0.5&1.25 &second& order& phase &transition  \\
~0.5&  2  &second& order& phase &transition \\ \hline
~\\~
      Table 3.   
\end{tabular}

The second column shows the values of the parameter $K = m^2_H/M^2_w$, 
corresponding to the Higgs boson masses 75 GeV, 90 GeV and 115 GeV, 
respectively. All the parameters are as in section 8.
   
It is  seen, an increase in $h$ makes the
 first order phase transition  weaker,  as in the case of hypermagnetic field.  For
strong fields $H \ge 10^{22} - 10^{23}$ G the baryogenesis condition $R \ge 1.2 - 1.5$
\cite{Shap2} is not satisfied.

Now, let us show the results for the asymptotic EP. They are gethered in Table
4. 

~\\~
\begin{tabular}{c c c c c c}\hline
~\\~
~h& $K$ &  $T_c(GeV)$& $\phi_c(h,T_c)$& $R$ & $M_w(h, T_c)$ \\[5pt] 
\hline
~0.01& 0.85 &106.321&0.0316 &0.0732 &0.1310  \\
~0.01& 1.25 &122.022&0.0548 &0.1104 & 0.1517  \\
~0.01&  2   &145.959&0.0616 &0.11   &0.1640 \\ \hline

~0.1 & 0.85 &107.267&0.0316 &0.073 & 0.1319     \\
~0.1 & 1.25 & second order& phase transition    \\
~0.1 &  2   & second order& phase transition   \\ \hline
~\\~
Table 4.  
\end{tabular}

First of all notice, the baryogenesis condition $R \ge 1.2$ is not satisfied 
for neither the exact EP nor the asymptotic one.
The $R$ values are small for for all $K$ considered.
 Since  the
external field is accounted of exactly through the Green
functions the effects of the field influence are also exactly reproduced. Thus, our
analysis has shown
that in the case of external magnetic field baryogenesis can not be generated in
the SM. The comparison of the results for the exact EP (Table 3) and the asymptotic
EP (Table 4) shows that in the latter case the second order phase transition is
predicted for more weak external fields.

The main results of this and previous section can be summarized as follows:
the  external either the  magnetic or the hypermagnetic field  can not produce the strong
first order EW phase transition. Baryogenesis does not survive in the SM due to these
external conditions.

In the next two sections we shall investigate the problems dealing with the
spontaneous magnetization of the vacuum at high temperature and stabilization of this
state due to radiation corrections. We shall show that strong magnetic fields $H
\sim g^{4/3}T^2$ have to be spontaneously created.

\subsection*{10. RADIATION SPECTRA OF $W$-BOSON IN STRONG MAGNETIC FIELD AT HIGH
        TEMPERATURE }

 $W$-boson mass operator  in a constant magnetic field $H$ at zero temperature has
        been calculated and investigated in Refs. \cite{SRV}, \cite{Skav}. In
 particular, it gave possibility to clarify the role of the radiation corrections
in the problem of  stabilization of the $W$-boson spectrum  (see survey
\cite{Skav}). The temperature dependent  radiation corrections  to the 
$W$-boson spectrum has  been studied in Ref. \cite{SS}. The longitudinal
components of gauge fields  acquire the temperature masses $\sim g T $ \cite{Kal}. The
tachyonic mode is the transversal state excited by a spin interaction.
 Its tepmperature mass can be calculated as the average value of
the mass operator in the ground state of the spectrum.

To incorporate temperature the imaginary time formalism   will be used. As in the case of
 zero temperature \cite{SRV}, the Schwinger operator method and
$s$-representation will be
 applied. In general, after summation over discrete imaginary frequencies this 
 method becomes not applicable. However, it remains practically unchanged in
 the limit of high temperature when only the static modes  $l=0 $ contribute
 \cite{Kal}. This approximation  is sufficient to investigate the role of the 
 daisy diagrams and will be used in what follows. 

We calculate the average value of the $W$-boson mass operator in the states of
 $W$-boson spectrum in a magnetic field $\mid n,\sigma>$, where $n,
 \sigma$ are the Landau level number and the spin projection variable,
 respectively. These functions, $< M(H, T, n , \sigma ) >$,  give the
 radiation temperature dependent masses of the states. The effective 
 mass squared is $M^2(H,T) = M^2 - eH + Re < M(H,T,  n = 0, \sigma
  = 1) >$. If this value is positive, the spectrum  is stabilized by
 radiation corrections.

In the present section we consider a simplified model of electroweak interactions (the
        boson part of it) based on the spontaneous breaking of $SU(2) \rightarrow 
        U(1)$ gauge group. The Lagrangian is
\begin{equation} \label{Lag} 
L = -\frac{1}{4} F^2_{\mu \nu} + \frac{1}{2} (D_{\mu} \phi)^2 + \frac{m_0^2}{2}\phi^2
       - \frac{\lambda}{4} \phi^4 ,
\end{equation}
where $x^2 = x^a x^a, a = 1, 2, 3 $, $ F^a_{\mu\nu} =  \partial _{\mu} A^a_{\nu} -
        \partial_{\nu} A^a_{\mu} + g \epsilon^{abc} A^b_{\mu} A^c_{\nu}$, $
        D_{\mu}^{ab} = \partial _{\mu} \delta^{ab} + g \epsilon^{abc}A^c_{\mu}$.
        This model is described, for example, in  Ref. \cite{Skav}. After the spontaneous
        symmetry breaking the charged components $W^{\pm}_{\mu} =
        \frac{1}{\sqrt{2}} (A^1_{\mu} \pm i A^2_{\mu} )$ asquir the masses $M = g
        \phi_c$, where $\phi_c = \mid m_0 \mid /\sqrt{\lambda}$ is the vacuum value of the
        scalar field, and the component $A_{\mu} = A^3_{\mu}$ remains massless
        and is identified with the electromagnetic potential. We also identify the gauge
        coupling constant $g$ with electric charge: $g = e$. In fact, this model is the
        mass regularization of the Yang-Mills theory. In the limit $M \rightarrow 0
        $ all the corresponding results are reproduced.  The results for
Salam-Weinberg theory
        can also be obtained by the simple  substitutions, $\gamma $-contributions
        $\rightarrow Z$-contributions with the corresponding vertex and propagator factors 
        (see for details Ref. \cite{Skav}).

To investigate the problem we are interested in, let us, as before, direct the
external
magnetic field along the
 third axis $H = H_3.$ The corresponding  potential is chosen in the
 form $A^{ext}_{\mu}$ = $(0, 0, H x^1, 0)$, $ H = const$. To quantize the
           fields the following gauge fixing condition  is used:
\begin{equation} \label{gauge}
\partial _{\mu} W^{\pm\mu} - ie A^{ext}_{\mu} W^{\pm\mu} - M
           \phi{\pm} = 0,
\end{equation}
where $\phi^{\pm} = \frac{1}{\sqrt{2}} (\phi^1 \pm i \phi^2 )$ are the charged
           Goldstone fields.  

The $W$-boson mass operator in one-loop order is given by a standard series of
diagrams \cite{Skav}, \cite{SRV}.  It can be written in the form
\begin{equation} \label{oper}
M_{\mu \nu} = \frac{e^2}{\beta} \sum \limits_{k_4 } \int \frac{d^3 k}{(2\pi)^3} [
           M^{\phi}_{\mu \nu} (k, P) + M^{vec.}_{\mu \nu} (k,P)],
\end{equation}
where $\beta = 1/T$, $k_4 = 2\pi l /\beta,~ l = 0, \pm 1, \pm 2, ...$, operator 
$  P_{\mu} = i\partial _{\mu} + e A^{ext}_{\mu}$. The first term in the Eq.
(\ref{oper}) describes  the contribution of virtual  neutral scalar  particles and 
the second one gives that of virtual gauge fields (photons, W- and Goldstone bosons).

 We restrict ourselves by investigation of the high temperature
        limit, that corresponds to the static mode $l = 0$.    In this case the standard
calculation procedure developed at zero
        temperature can be straightforwardly applied. Details of  these calculations can
        be found in Refs. \cite{SRV}, \cite{Skav}, \cite{SS}. Here, we note the points
specific
        for $k_4 = 0$ case. 
For brevity, in what follows we will consider the part $M^{\phi}(k,P).$

To integrate over $d^3 k$ in Eq. (\ref{oper}) one has to introduce the  
s-representation for each propagator entering and present the product of 
        propagators as follows
\begin{equation} \label{repr} D_{\phi}(k) G_{w}(P - k) = - \int \limits_0^1 d u \int
        \limits_0^{\infty} d s  s e^{is \cal{H}},
\end{equation}
where "Hamiltonian" is,
\begin{equation}
 {\cal H} = (1 - u) \vec{k}^2 + u ( \vec{P} - \vec{k})^2 - u (M^2 + 2ieF) -
        (1 - u )m ^2,
\end{equation}
\nonumber
and  $m$ is the mass of the neutral scalar field. The $k_4$-dependent part of 
the Hamiltonian  is  zero. Then, the integration is carried out by means of the 
 procedure developed in Ref. \cite{Schw}. The difference between the finite and zero
temperature
 cases consists in the dimension of the corresponding integrals. This is reflected 
in the power of $s$ appeared after integration. Three dimensional integral (at $T 
\neq 0$ ) gives the factor $s^{-3/2}$ whereas at $T = 0$ one obtains  $s^{-2}$
\cite{Schw}.

To find the energy of the states owing to the radiation corrections one has to define the
        $W$-boson mass-shell in the field at high temperature. In the static case it is
        described by the equations
\begin{eqnarray} \label{msh}
[&(&\vec{P}^2 + M^2 ) \delta_{ij}  + 2ie F^{ext}_{ij}~ ]~ W^{-}_j = 0 , \\ \nonumber
&&P_i W^{-}_i = 0 ,~~   \phi^{-} = 0 ,
\end{eqnarray}
where $j = 1,2,3$ and the product $P_4 W_4 = 0$ bacause for static modes $p_4 = 0$. 
 The states are normalized by the condition
\begin{equation} \label{nor}
< n, \sigma \mid  n', \sigma' > = \delta_{n,n'}~ \delta_{\sigma,\sigma'},
\end{equation}
$n, n' = 0,1 , ...$ and $\sigma, \sigma' = 0, \pm 1$. 

The average value of the mass operator in these states can be written in the form
\begin{eqnarray} \label{mass}
< M >& =& \frac{\alpha}{2\sqrt{\pi}\beta} \int\limits_0^1 d u \int\limits _0^{\infty}
        \frac{d x}{\sqrt{x}} [e u H \Delta ]^{-1/2} e^{-x u M^2/eH} \\ \nonumber
&exp&[-(2n + 1) ( \rho - x(1 - u) - 2 u (1 - u) ] M( x, u ) ,
\end{eqnarray}
where $ M( x, u ) = < n, \sigma \mid M_{ij} \mid n, \sigma >$ and
\begin{eqnarray}
tanh\rho &=& \frac{(1 - u ) sh x }{ (1 - u )ch x + u sh x / x} \\ \nonumber
\Delta &=& (1 - u )^2 + 2 u (1 - u ) \frac{sh 2x}{2x} + u^2 (\frac{sh x}{x})^2.
\end{eqnarray}
\nonumber 
This very complicate expression can be investigated for different limits of interest. In
        particular,  for high temperatures and strong fields, $\frac{eH}{M^2}>> 1,
        \frac{eH}{T^2} << 1$ we obtain (in the reference frame $p_3 = 0$):
\begin{eqnarray} \label{corr}
< n, \sigma = + 1\mid M^{gauge} \mid n, \sigma = + 1> &=& \frac{e^2}{4\pi}
        (eH)^{1/2} T [ 12,33 + 4 n \\ \nonumber
 &+& i ( 3 + 6 n ) ] ,\\ \nonumber
< n, \sigma = + 1 \mid M^{\phi} \mid n, \sigma = + 1 > &=& \frac{e^2}{4\pi} (eH)^{1/2}
        T [ 14,63 + 4 n + i ( 7 + 6 n )], \\ \nonumber
Re < n, \sigma = - 1 \mid M^{gauge} \mid n, \sigma - 1 > &=& \frac{e^2}{4\pi}
        (eH)^{1/2} T ( 11,44 + 4n ), \\ \nonumber
Re <n, \sigma = 0 \mid M^{gauge} \mid n, \sigma = 0 >  &=& \frac{e^2}{4\pi}
        (eH)^{1/2} T ( 15,44 + 4n ).
\end{eqnarray}
Above formulae give a general picture on the
 behaviour of the radiation energies for different spin states. As it is seen, the real
 part of $< M >$ is positive in the ground and excited states. It acts to
 stabilize the tree spectrum. The imaginary part describes the decay of the states
 due to transitions to the ones having lower energies. 

Now, let us consider the effective $W$-boson mass squared:
\begin{eqnarray} \label{Meff}
M^2(H,T) &=& M^2 - eH \\ \nonumber
 &+& Re < n = 0, \sigma = + 1 \mid M^{gauge} + M^{\phi} \mid
        n = 0, \sigma = + 1 > \\ \nonumber
&=& M^2 - eH + 26,96 \frac{e^2}{4 \pi} (eH)^{1/2} T.
\end{eqnarray}
This value is positive for sufficiently high temperatures. Hence, one can conclude that
 radiation corrections in the field stabilize the vacuum at high temperatures.  The
temperature mass of the transversal modes dependens on the field 
 and equals to zero when $H = 0$, as it should be \cite{Kal}.

Obtained results are of interest for cosmology. Namely, if  at the EW phase
 transition epoch the magnetic field was present,  the radiation
 mass  of $W$-bosons could serve as the dynamic mechanism of the vacuum stabilization
 in both the broken and the restored phases. It was discussed in detail in the previous
sections.  To derive  a consisten picture one has to
        consider a vacuum magnetization at high temperature with the correlation
        corrections taken into consideration. 

\subsection*{ 11. THE SPONTANEOUS VACUUM MAGNETIZATION AT HIGH 
        TEMPERATURE }

 The generation of magnetic fields in nonabelian gauge theories at finite temperature
 is of great importance for particle physiscs and cosmology. Its positive solution, in
 particular, will give a theoretical basis for investigations of the QCD vacuum at
  high temperature and the primordial magnetic fields in the early universe
        \cite{Enol}, \cite{Amol}. In literature different mechanisms of producing the
        fields are discussed (see recent papers \cite{Shap1}, \cite{Shap2} and
references
        theirin). In this section we  investigate in more detail one of possibilities
- the spontaneous magnetization of the
        vacuum of nonabelian gauge fields at finite temperature. This problem has been
        studied in one-loop order in Refs. \cite{Enol}, \cite{SVZ1}, \cite{Ska3} where
        the creation of the vacuum field was derived. In Ref. \cite{SVZ1} a number of
        correlation corrections has been taken into account, but the polarization
        functions of gauge fields were not calculated and therefore a
 trusty
        conclusion about the phenomenon was not obtained. In Ref. \cite{Bors2} that
has been considered with all necessary daisy diagrams included. Below, we shall follow
 this paper. 

For simplicity, we investigate the vacuum magnetization at finite temperature within
        $SU(2)$ gluodynamics. Considering the abelian covariatly constant
        chromomagnetic field $H^a = \delta^{a3}H = const$ and finite temperature as
        a background, we calculate the EP containing the one-loop and  daisy
        diagram contributions of the neutral and charged gluon fields.
The field dependent Debye mass of neutral gluons can be computed from a special type EP.
 To find the one of the charged gluons  the high temperature limits of the gluon
polarization functions   at the background are also calculated. It will be shown that in
the
        adopted approximation the Savvidy level with the field strenght $(gH)^{1/2}_c
        \sim g^{4/3} T$ is generated. This is strong field. 
Although it is screened at distances $l > (g^2 T)^{-1}$
by the gluon
        magnetic mass, the spectrum of charged particles, being formed at Larmor's
        radius $r \sim (gH)^{-1/2}_c \sim (g^{4/3} T)^{-1}$, is located inside this
domain, $ r << l_{m}$, for small $g$.

Most of results for the Yang-Mills (YM) theory can be obtained without actual
 calculations  by making use  of the results of the previous chapters. Really,  the model
considered in sect. 9 is a mass regularization of the YM theory. The external
magnetic field just coinsides with the abelian  chromomagnetic field of interest.
Therefore, the results for the latter one
 can be
        obtained by setting the mass of the charged gauge fields to zero and
        subtracting the contributions of  the longitudinal spin projection of the massive
        charged gauge fields and the Goldstone fields. These 
        transformations can be  made freely.  For instance,  to get the
        contribution of charged gluons to the one-loop EP it is sufficient to substitute
        the factor 3 by the factor 2 in the part of Eq. (\ref{13})  describing the
        contribution of massless ($M = 0$) charged spin zero particles. The same has
        to be done 
to find either the Debye mass of neutral gluons (in Eq. (\ref{31})) or the
 temperature  masses of charged gluons  (sect. 9), etc. Below, for convenience of
 further account, we adduce the necessary  expressions \cite{Bors2}.

 The high temperature limit of the one-loop EP is 
\begin{eqnarray} \label{V1limY}
 V^{(1)}(H,T)&=& \frac{H^2}{2} + \frac{11}{48}\frac{g^2}{\pi^2}H^2
        log\frac{T^2}{\mu^2} - \frac{1}{3}\frac{(gH)^{3/2}T}{\pi}\nonumber\\ 
&-&i\frac{(gH)^{3/2}T}{2\pi} + O(g^2H^2).
\end{eqnarray}
Notice the  cancellation of $H$-dependent logarithms entering the vacuum and
        the statistical parts, $\mu$ is a normalization point.

\subsection*{11.1 Daisy diagrams of the unstable mode and the neutral gluons}

The contribution of daisy diagrams with the unstable mode is given by the expression
 \cite{Bors2}
\begin{equation} \label{VunstY}
V_{unstable} = \frac{gH T}{2\pi} [\Pi(H,T,n=0,\sigma=+1) - gH ]^{1/2} +
        i\frac{(gH)^{3/2}T}{2\pi}.
\end{equation}
From Eqs. (\ref{V1limY}) and (\ref{VunstY}) it is seen that the imaginary terms are
        cancelled out in the total. The final EP is real if the condition 
        $\Pi_{unstable}(H, T) > gH $ holds. 

This expression must be supplemented by the term describing the contribution of
the neutral
        gluon fields. In one-loop order  it gives a trivial $H$-independent constant
        which can be omitted. However, these fields are long-range states and they do
        give $H$-dependent EP through the correlation corrections including  the mass
        term $\Pi^{0}( H,T)$.  Corresponding part  of the EP is described by the
        expression which can be recognized from  the case $H = 0$  \cite{Car},
        \cite{Kal}:
 \begin{eqnarray} \label{Vring1Y} V_{ring} &=&
\frac{1}{24}\Pi^{0}(H,T) T^2 
 -\frac{1}{12\pi\beta} [\Pi^{0}(H,T)]^{3/2}\nonumber \\ 
&+& \frac{(\Pi^0(H,T))^2}{32\pi^2}[log(\frac{4\pi T}{(\Pi^0)^{1/2}}) + \frac{3}{4} - C ],
\end{eqnarray}
where $\Pi^0(H,T)= \Pi^0_{00}(k=0,H,T) $ is the zero-zero component of the neutral
        gluon  polarization operator calculated in the external field at finite
        temperature and taken at zero momentum, C is Euler's constant. The
        first term in Eq. (\ref{Vring1Y}) has order $\sim g^2$ in coupling
constant,
        the second term is of order $\sim g^3$ and the last one - $\sim g^4$.  
Restricting ourselves by order $\sim g^3,$ it will be
        omitted in what follows.  As usually, for $\Pi(H,T)$ the high temperature limit
        of the function has to be used. The expression (\ref{Vring1Y}) needs in one
        additional comment. If one compares it with Eq. (\ref{21}), the 
        difference will be in the extra term  $\sim \Pi(T,H) T^2$.
 We have maintained this next-to-leading term because it is of great
        importance for problems under consideration. In studying of symmetry
        behaviour only the leading ($\phi$-independent) terms of the polarization
        functions were taking into account.

The mass squared $m_{D~neutral}^2 = \Pi_{00}(H,T, p_0 = 0)$, that is the Debye mass
        of neutral gluons, reads
\begin{equation} \label{md1} m^2_D =  \frac{2}{3}g^2 T^2 -
\frac{(g H)^{1/2}}{\pi}T - \frac{1}{4\pi^2}(g H)
+ O((gH)^2/T^2).
\end{equation}
Here, the first term is the well known temperature mass squared and other ones
        give the field-dependent contributions. They have negative signs that is
        important for what follows. Sabstituting expression (\ref{md1}) into equation
        (\ref{Vring1Y}) we obtain the correlation corrections due to neutral gluons. 

The calculations, as we have described in sect 10., result in the high temperature limits
of
        $\Pi_{unstable}(H,T)$
 \begin{eqnarray} \label{munst} \Pi_{unstable}(H,
T) &=& <n=0,\sigma= + 1\mid \Pi^{charged}_{\mu \nu} \mid n=0, \sigma = + 1>
\nonumber\\
 &=&
12.33 \frac{g^2}{4\pi} (gH)^{1/2} T , 
\end{eqnarray}
 and of excited states $ \Pi(n \ne 0, \sigma)$,
 \begin{eqnarray}\label{mst}
 Re \Pi(p_4 = 0,n, p_3 = 0, H,T, \sigma = + 1) =
\frac{g^2}{4\pi} (gH)^{1/2} T ( 12.33 + 4 n ), \\ \nonumber 
Re\Pi(p_4 = 0,n, p_3 = 0, H,T, \sigma = - 1) = \frac{g^2}{4\pi}
(gH)^{1/2} T ( 11.44 + 4 n ),
 \end{eqnarray}
where the average values of the polarization operator  in the corresponding states of the
        spectrum (\ref{spectr}) are computed. These formulae have been obtained in  
the  high temperature limit $gH/ T^2 << 1$.  The operator contains also an imaginary
part which describes the decay of the states.
But for the problem
 under consideration only the real part is needed because it is responsible for the
radiation masses of particles.

Let us note the most important features of the expressions (\ref{munst}), (\ref{mst}). 
It is  seen, at $H = 0$ no screening magnetic mass is produced in one-loop order.
Second, the mass squared of the modes are positive and
        act to stabilize the spectrum of charged gluons at high temperatures. Therefore,
in the nonzero chromomagnetic field at finite temperature the
        charged transversal gluons become massive.

\subsection*{11.2  Contribution of the  daisy diagrams  of charged gluons}

To obtain the correlation corrections due to the logitudinal charged gluons the
zero-zero
 component of
        the polarization operator has to be calculated. The Debye mass of charged
        gluons is found to be \cite{SS}
\begin{equation} \label{dbmch}
\Pi_{00}(k_4 = 0, k_3 = 0, H,T) = \frac{2}{3} g^2 T^2 +  \frac{g^2}{4\pi} (gH)^{1/2} T
        (6 + 4 n),
\end{equation}
where  again only the real part is adduced. Hence, the next-to-leading terms are
  the growing positive functions of $n$. 

Now, let us turn to the generalized EP
\begin{equation} \label{EPtY}
V^{(1)}_{gen} = \frac{gH}{2\pi}\sum \limits_{l= -\infty}^{+ \infty} \int\limits_{-
\infty}^{+ \infty} \frac{dp_3}{2\pi} \sum\limits_{n = 0, \sigma =
\pm 1}^{\infty} log [\beta^2(\omega^2_l + \epsilon ^2_{n,\sigma,p_3}
+ \Pi(n,T,H,\sigma) )] 
 \end{equation}
 written as the sum of energies of the charged gluon field modes in the external
       chromomagnetic field ( $ \omega_l = \frac{2\pi l}{\beta}$ - discrete imaginary
       energies)
\begin{equation} \label{spectr} \epsilon^2_{n,\sigma} = p^2_3 + (2n +
1 - 2\sigma) gH + \Pi(H, n,\sigma, T) 
 \end{equation}
and including  the temperature masses $\Pi(H, n,\sigma, T)$ of the modes which are also
       dependent on $H$, the level number, n, and the spin projection $\sigma$. 
In general, the incorporation of the polarization functions into EP may result in wrong
       combinatoric factors for the two-loop diagrams. In our case, to be sure in the
       obtaining results the following procedure is applied. We subtract the term
\begin{equation} \label{loop}
V_s = \frac{gH}{2\pi} \sum\limits_{l=-\infty}^{+\infty} \int\limits_
{-\infty}^{+\infty}\frac{dp_3}{2\pi} \sum\limits_{n=0,\sigma = \pm 1}^
{\infty} D_0(p_3, H, T) \Pi( H, n, T, \sigma)
\end{equation}
from Eq. (\ref{EPtY}), that separates the contributions of  the two-loop diagrams of
 charged gluons. One freely can check that the two-loop diagrams containing the one of
 loops with the neutral gluon lines can be accounted of within the generalized EP
 including the insertion of the neutral gluon
       polarization operator. After that the  only two-loop diagram (two touch
circles) allowing for
       the self-interaction of charged gluon fields in the vacuum  remains to  be
computed
       separately.

Then, by substituting the expressions (\ref{munst}), (\ref{mst}), (\ref{dbmch}) into
        Eq. (\ref{EPtY}) and integrating over momentum and calculating the sums in
        $n$, we obtain the daisy diagram contribution of charged gluon fields. The
        result can be expressed in terms of the generalized $\zeta$-function and looks
        as follows,
\begin{eqnarray} \label{vrch}
V_{ring}^{ch}& = &\frac{gH T}{2\pi} \big \{ \sqrt{2 gH_D}[ \zeta 
( - \frac{1}{2}, a_+ ) 
+ \zeta ( - \frac{1}{2}, a_- ) 
+ 2 \zeta 
( - \frac{1}{2}, a_D)] \nonumber \\
& -  &\sqrt{2 gH} [3 \zeta (-\frac{1}{2}, \frac{1}{2} ) + 
\zeta (- \frac{1}{2}, \frac{3}{2})] \nonumber \\
&+& (\Pi(H,T,n=0,
\sigma=+1)^{1/2}\big\},
 \end{eqnarray} 
where the first term in the first squared brackets corresponds to the spin projection
 $\sigma = + 1$, the second term - $\sigma = - 1$ and the last one describes the part
        due to longitudinal charged gluons. The terms in the second squared brackets
        give the independent of $\Pi(H,T)$ part of eq. (\ref{EPtY}). The last term in
        the curly brackets is due to radiation mass of the unstable mode.
  The notations are introduced:
 $gH_D = gH +
\frac{g^2}{2 \pi} (gH)^{1/2} T $ , $ a_- = \frac{1}{2} + \frac{g^2}{4\pi}
\frac{11,44(gH)^{1/2} T}{2 gH_D}, a_+ = \frac{1}{2} + \frac{g^2}{4
\pi}\frac{19,62 (gH)^{1/2} T}{2 gH_D}$ and $ a_D = (\frac{2}{3} g^2 T^2 + 
\frac{3g^2}{2\pi}(gH)^{1/2}T + gH )/2 gH_D $. 
 This expression is real for sufficiently high temperatures.

Substituting the expression (\ref{md1}) in eq. (\ref{Vring1Y}) and gethering
 all other contributions, we obtain the  consistent EP. 

To calculate the two-loop vacuum diagram describing the self-interaction of 
 charged gluons the standard procedure can be applied.
 The computation details are given
       in Ref. \cite{Bors2}. Reffering the readers to this paper, we write down here the
       resulting expression for the high temperature limit 
\begin{equation} \label{V2chlim}
V_{ch}^{(2)}(H,T)_{\mid _{T \to \infty}} = - \frac{g^2}{4 \pi}
(gH)^{1/2} T^3 + O(gH T^2).
\end{equation}
The high temperature limit of the term subtracted from the generalized EP 
 (\ref{loop}) is
\begin{equation} \label{loopl}
V_s(H,T)_{\mid _{T \rightarrow \infty}} = \frac{g^2}{6\sqrt{2}\pi} \zeta(
\frac{1}{2},- \frac{1}{2}) (gH)^{1/2} T^3 + O(gH T^2).
\end{equation} It also gives the negative 
        contribution to the leading terms of the asymptotic expansion since $\zeta
        (\frac{1}{2},-\frac{1}{2}) = 0.8093$. It worth to mention that these are the
        longitudinal modes that determine the high temperature behaviour of
        $V^{(2)}_{ch}(H,T)$. Having obtained the two-loop corrections to the EP, one
can 
investigate the spontaneous vacuum magnetization at high temperature. 

\subsection*{11.3 Vacuum magnetization and the stability problem}

The derived EP is expressed through the well studied special functions. Therefore, it can
        easily be investigated numerically for any range of parameters entering. As
    usually, it is convenient to introduce the dimensionless variables: the field $\phi
        = (gH)^{1/2}/T$ and the EP $v(\phi,g) = V(H,T)/T^4$. The vacuum
        magnetization at high temperatures, $ T>> (gH)^{1/2},\phi \rightarrow 0$, will
        be investigated within the followin limiting form of the EP,
\begin{eqnarray} \label{Vlim} v^{total}(\phi,g)_{\mid \phi
\rightarrow 0} &=& \frac{\phi^4}{2 g^2} +\frac{11}{48} \frac{\phi^4}{\pi^2} 
log(\frac{T^2}{\mu^2})  - \frac{1}{3} \frac{\phi^{3}}{\pi}
- \frac{g^2}{48\pi} \phi
\nonumber\\ 
&-&\frac{1}{3} (\frac{2}{3})^{3/2}\frac{g^3}{\pi}
\frac{\phi + \frac{27}{8}\frac{\phi^2}{\pi}}{\phi + \frac{g^2}{2\pi}}
- \frac{g^2}{4\pi} \phi - \frac{g^2}{6\sqrt{2}\pi} \cdot 0,8093 \phi,
\end{eqnarray}
where other $\sim g^3$ terms are omitted.
The logarithmic term is signalling the asymptotic freedom of $g^2(T)$ at high
        temperatures \cite{Ska3}. It includes explicitely the dependence on the scale
        parameter $\mu$. Other terms present, respectively, the high temperature limits
        of the one-loop EP,  the neutral gluon and the charged gluon daisyes and the
        two-loop diagram of charged gluons. To obtain the term due to
        $V_{ch}^{ring}$ the asymptotic expression for Zeta-function \cite{Odin},
\begin{equation} \zeta(-\frac{1}{2}, a_D)_{\mid a_D \rightarrow \infty} =
 - \frac{2}{3} a_D^{3/2} + \frac{1}{2} a_D^{1/2} - \frac{1}{48} a_D^{-1/2} 
+ O(a_D^{-3/2})
\end{equation}\nonumber
was used. Zeta-functions with $a_+, a_-$ do not contribute in leading order. Since we are
        searching for  the fields $\phi$ of order larger then $g^2$,  we can omit the
        term $g^3$ and obtain for the condensed field
\begin{equation} \label{mf} (gH)^{1/2}_c = \frac{0.6}{\pi^{1/3}}\frac{}{}g^{4/3} T .
\end{equation}
Thus, we come to the result that the ferromagnetic vacuum state indeed exists at high
        temperatures. The correlation corrections increase the field strength as compare
        to the one-loop value ( $(gH)^{1/2}_c \sim g^2 T)$. 

Let us discuss the stability of the condensed field. The second derivative of the
EP is positive for $H_c$ that means we have the minimum. The field is not changing
in the direction $a = 3$ of the isotopic space. To check that this is indeed the
case for the perpendicular directions $ = 1, 2$ or $a = a^{\pm}$ responsible for
excitation of charged fields $W^{\pm}$, one has to calculate the effective mass
 squared $M^2(H_c, T)$. First we consider the one-loop case. Sabstituting the
value $(gH_c^{(1)})^{1/2} = (g^2/2\pi) T$ in the one-loop polarization function,
we find that the effective mass squared, $M^2(H_c^{(1)}, T) = \Pi^{(1)}(H_c^{(1)},
n = 0, \sigma = + 1) - gH_c^{(1)} \ge 0$, is positive. Thus, the vacuum
stabilization is observed in this consisten calculation. However, if one checks
whether the one-loop gluon radiation mass stabilize the true vacuum magnetic field
and substitutes the value $(gH_c)^{1/2} \sim g^{4/3} T,$ the negative value of
$M^2(H_c, T)$ will be obtained. The one-loop mass does not stabilize the spectrum
and, hence, vacuum. Nevertheless, the one-loop result makes hopeful the idea to
have the stable vacuum due to radiation corrections to the charged gluon spectrum.
Naturally, to investigate this possibility the gluon polarization operator with
the correlation corrections included should be calculated. This problem requirs an
an additional investigation. Other interesting possibility is the formation at
        high temperatures of the gluon electrostatic potential, so-called $A_0$
        condensate (see survey \cite{BBS}), which also acts as a stabilizing factor
        \cite{SVZ1}. To realize the latter scenario consistently the simultaneous 
        spontaneous generation of both the $A_0$ condensate and the magnetic field
   should be investigated. If again the
        homogeneous vacuum field will be found to be  unstable with these 
        improvements made, the inhomogeneous fields of the lattice type discussed in
        Refs. \cite{Amol}, \cite{Sk1}, \cite{MacD} can be created. Really, since the
        condensed magnetic field is strong at high temperatures, the lattice structures
        having the cells of order $\sim 1/ (g^{4/3}T) << 1/(g^2 T )$ are located inside
        the domain where the fields are not screened by the gluon magnetic mass. 

 The  above calculations have unambiguously determined the possibility of the vacuum
magnetization at high temperature, although a number of questions
concerning the vacuum stabilization and  structure has to be investigated in order to
derive a final picture. This could result in the presence of strong magnetic fields in
the hot universe.

\subsection*{12. COMPARISON  WITH OTHER APPROACHES}

In this section we are going to compare our results  with  that of other
investigations. That refer to either  the phase transition  or the generation
of the magnetic field at high  temperature.  We begin with discussion of the EW
phase transition in strong hypermagnetic fields.

First let us discuss the results  of Refs. \cite{Shap1}, \cite{Elm1}  
where it was concluded that strong external hypermagnetic field generates the
sufficiently strong first order phase transition and baryogenesis 
survives in the SM. In these papers the inflence of the field on the  vacuum
has been allowed for in tree approximation that gives
the qualitative estimate of the effect.  In Ref. \cite{Elm1} the temperature
dependent
part of the EP has been included in one-loop order whereas the field
dependence was skipped.  Naturally, 
further studying of the  phenomenon should include the field dependent radiation and
correlation corrections due to fermions and bosons.  This is the problem that we
have addressed to in the present investigation. The main idea was to
determine the form of the EP curve in the broken phase and find the  
range of the parameters $H_Y, K $ when  the EW phase transition of first
order is happened.  To elaborate that the consistent EP including the
one-loop and daisy diagrams of all the fundamental particles has been
calculated. As we have discovered, the role of fermions is crucial in a vacuum
dynamics  in strong fields at high temperature. They essentially affect the
structure of the broken phase making the EW phase transition weaker as compare to
the tree level results. The external field has been accounted of exactly through
Green's functions. The minimum of the EP was found to be real at
sufficently high temperatures when the first order phase transition
happens. This important property was established when the daisy diagrams of the
tachyonic mode have been included. As a result, no conditions for $W$- and
$Z$-boson condensates are observed at high temperatures and the external field
strengths
corresponding to  the first order phase transition. The condensates could be
generated for stronger fields when the phase transition  becomes of second order. 
But, this is not of interest here since we are looking for conditions when  
baryogenesis can survive.

In Refs. \cite{Shap4}, \cite{Lain} the EW phase transition in the
hypermagnetic field has been investigated by
 means of a general method developed in Ref. \cite{Kaja} (see also survey
  \cite{Shap2}) which combines the perturbative computations and the lattice simulations. 
 The results obtained therein have supported the main conclusions of Refs.
\cite{Shap1},
 \cite{Elm1} discussed above. In more detail,  it was found for the fields $H_Y <
0.3 T^2$ the first order phase transition  becomes stronger, but it still turns
 into a crossover for masses $m_H \geq 80 $ GeV. For stronger fields, a mixed phase
 analogous to a first type  superconductor with a single macroscopic tube of symmetric
 phase, parallel to $H_Y$, penetrating through the broken phase, has been observed.
  
As this aproach is concerned, we note that because of peculiarities of the
calculation
procedure adopted in Ref. \cite{Shap4} all fermions except t-quarks are  decoupled as
nonstatic modes and therefore the field dependent fermion contributions  as well as
the correlation corrections could not be accounted of.
That is why these calculations also do not reproduce correctly symmetry behaviour  for
 strong external fields. 

In the present approach, the value of the mass $m_H = 75 $ GeV ( = 0.85)  is close to
$m_H = 80$ GeV discussed in Ref. \cite{Shap4}. This is important for us
 because the first order phase transition is controlled by perturbative 
method used. But we could not observe the mixed phase. We  have seen the
crossover (or second order phase transition) for $H_Y = 10^{23}$ G and $K$ = 0.85
determined within the asymptotic EP. The exact EP in this case predicts the
weak first-order phase transition. Since the EP is real in the minimum, no conditions
for the vortex-like phase exist. 

The Higgs boson mass values considered correspond to  the cases when perturbative
results are reliable ($K$ = 0.85) and may be 
not trusty (K = 1.25 , 2). However, since the external field is taken
into account exactly its effects are correctly reproduced. As we have seen, an
increase in $H_Y$ makes the EW phase
transition of second order for field strengths $H_Y \sim 0.5 \cdot   
10^{24}G$ for all the values of $K$ investigated. For weaker fields the
phase transition is of first order but the ratio $R = \phi_c(H,T_c)/T_c$
is less then unit, that is unsufficient to generate baryogenesis.  

Let us remind the situation  with the magnetic field when the magnetic mass
 of order $m_{mag} \sim g^2 T$ is taken into consideration.  This mass screens
the nonabelian
 component $H \delta^{a3}$ at distancies $l > l_m \sim (g^2 T)^{-1}$ but inside
 the space domain $l < l_m$  it may exist and  affect all the processes at high
temperatures. The latter fact has not been
realized in a number of investigations  \cite{Elmp},
\cite{Pers} where (as in Ref. \cite{Enol}) the
 field strength generated at finite temperature has been erroneously estimated as
 coinsiding with that at zero temperature. Our estimate of the field strength at
high temparature makes the investigation
of the EW phase transition in strong
magnetic fields resonable. But, as we have seen in sect. 9, in this case it  also is
impossible to generate the strong first order phase transition. 

Now, let us consider in more detail the problems of the magnetic field generation and
stabilization
 at high temperature. In the present survey we have investigated in detail 
 the Savvidy mechanism. The ways of the vacuum
stabilization have also been discussed. In this scenario, usual
 magnetic field should be treated as the projection of the chromomagnetic field
 created in a nonabelian gauge theory. Results of our investigation disagree
 with that of Refs. \cite{Elmp}, \cite{Pers} where it is claimed
 that spontaneous magnetization does not hold at high temperature.  To clarify
the origin of the  discrepancy let us repeat the main statements of Refs. \cite{Elmp},
\cite{Pers}: 1)  The field strength generated at finite
temperature  $gH \sim  \Lambda^2$ , where $\Lambda ^2 = \mu^2 exp(-\frac{48 \pi^2}{
11 N g^2(\mu)} ),$  coinsides with that at zero temperature. This is much
less than the  magnetic mass squared $ \sim (g^2 T)^2$. 2) Since the spectrum of charged
particles in the  magnetic field is formed at the Larmor radius scale, the weak long
range fields  are not produced being screened by the mass at distances $l \geq 1/ g^2 T.$
The error is the assumption that the field strength
generated in the vaccum is not changed when the  temperature is switched on.  As we
have  seen in sect. 11, strong abelian colour magnetic fields of order $(gH)^{1/2}
\sim g^2 T $ (in one-loop  approximation \cite{SVZ1}, \cite{Ska3}) or 
$(gH)^{1/2} \sim g^{4/3} T$ \cite{Bors2} (when  higher order corrections are
included) is  spontaneously generated at hight temperture.

We also have observed in the consistent calculation that the Savvidy state
is
stable at high
temperature when the  one-loop  EP and the one-loop gluon polarization operator are 
taken into account. 

\subsection*{13. DISCUSSION}

 As it was realized recently, in the
SM the usual scenario of  baryogenesis can not be established without external
 fields \cite{Shap2}. By analogy to superconductivity it was assumed that strong
hypermagnetic
field is able to generate the strong first order phase transition for the values of mass
$m_H$ permitted by experiment \cite{Shap1}. Different mechanisms to produce
magnetic fields have been proposed in literature. They can be
devided in two groups: 1) generation of fields at EW phase transition
\cite{Baym}, \cite{Cheng}, \cite{Kibble}, \cite{Ahonen}, \cite{Enqvist98};
2)
generation of fields beyond the SM scale (for instance, GUT
theories) \cite{Enq}, \cite{Enol}. A special interest was in strong
hypermagnetic field which due to its abalian nature
is not screened at high temperature whereas the nonabelian component of the usual
magnetic
field is screened at distances $l > 1/ g^2 T$ by the  gauge field magnetic mass.
The influence of the magnetic fields generated at the EW scale on the first order phase
transition was investigated in Ref. \cite{Fior}. It has been shown that if the
field strength is stronger then $H_0 = 10^{24}$ G the phase transition
is delayed.  Of cause, it is relevant if the first order phase transition is realized. 
But this is not the case for the SM, where without external fields for the mass $m_H
\ge 60 - 70$ GeV  the second order phase transition is predicted. Naturally, internal
forses are not able to change the kind of the phase transition.

The investigations carried out in the present paper are refered to the second 
possibility assuming that the fields have been generated before the EW phase
transition at a GUT scale. The dynamics of the hypermagnetic field generation has 
not been discussed here, although the corresponding machanism was proposed in
Ref. \cite{Shapg}. We just assumed that the field present in the early universe. As
the magnetic field is concerned, in sects. 10, 11 we have investigated in detail  its
generation and stabilization at high temperature.

The results on the hypermagnetic field influence can be summarised as follows. In
contrast to the conclusions of Refs. \cite{Elm1}, \cite{Shap4} claiming that strong
hypermagnetic  field  generates the strong first order EW phase transition, we
observed an opposite effect. As we have seen in our numerical
computations, the  weak first-order EW phase thansition bacomes of
second order in strong fields.
The origin of the discrepancy  is the following. These authors  have not determined
correctly the form of the EP curve at the transition temperature since  the 
contributions of fermions and the correlation corrections in the
fields were skipped. In fact, our analysis  has completed
the investigations not only for the hypermagnetic but also for the magnetic field.  We
have discovered that the role of fermion radiation corrections is of
great importance in the phenomenon investigated. 
 To better understand the role of
fermions in symmetry behaviour let us adduce two terms of the
asymptotic expansion of the EP in the limit of $T \rightarrow \infty, H \rightarrow
\infty$. The first one is the term
$\sim H^2 log \frac{T}{m_f}$. Due to this term the light fermions are
dominating at high temperature.
The second term can be derived from the expansion of the zero temperature part
Eq. (\ref{19}). This expression  side by side with the leading term $\sim H^2
log\frac{eH}{m_f}$, which due to a "dimension parameter trading" is replaced by the
above
written term, contains the subleading one $\sim - eH m_f^2 log\frac{eH}{m_f^2}$ (for
details see Ref. \cite{Ditt1}). This term acts to make "heavier" the Higgs particles
in the field.  As a result, the second order temperature phase transition  is
stimulated due to strong fields. We also have investigated the influence of different
parts of the EP on
symmetry behaviour. It was discovered that the change of the kind of the EW phase
transition  with increase in $H$ is due to the fermion temperature part of the
EP.

The stability problem of the vacuum at high
temperature and strong fields has also been investigated.  Our EP (its
minima) is real at the first order phase transition that is important for the
reliability of the results obtained. The latter property is insured by the 
imaginary terms of daisy diagrams cancelling the corresponding ones of the one-loop
EP. Thus, the total EP is sutable to correctly describe the phase transition. As we
have found,
the 
effective W-boson mass $M_w(H, T)$ is real in the local minimum, therefore no
conditions for the formation of the
vortex-like structure due to W- and Z-boson condensates exist in the broken phase.  
Probably, this is the reason why the condensates have not been observed in Ref.
\cite{Shap4}, although it is difficult to check that strightforwardly. 

Furthermore, we have also shown that spontaneous vacuum magnetization indeed happens at
high
temperature giving the  reliable mechanism of the magnetic field generation at
a GUT scale. In principle, the field could be spontaneously created at the
EW scale. However, in this case it is incorrect to treat the  EW phase transition as the
 external field problem. It is necessary to apply consideration as in Ref.
\cite{Fior}.    

On the base of our analysis one has to conclude that baryogenesis
can not survive in the SM  if  smooth external magnetic fields  generated beyond
the EW scale are included as environment. 

As a final remark, we would like to stress once again that light fermions act to
turn
the first order phase transition into the second order one with increase in the
field strength, independently of the values of the Higgs boson mass considered. 
This fact may be one of the reasons why it is of interest to carry out a  similar
 investigation for  the minimal supersymmertic SM  or other
supersymmetric extensions of it, because the  fermion and the boson sectors
enter such models on an equal footing.

The authors thank M. Bordag, A. D. Linde, D. Schildkneht and V. S. Vanyashin for  discussions and
remarks and A. Batrachenko for checking of the numeric calculations. One of us
(V.S.)
grateful Internationl Centre for Theoretical Physics
(Trieste, Italy)  and Institute for Theoretical Physics University of Leipzig
where a part of this work has been done for hospitality.
This work was partially supported by DFG  grant No 436 UKR 17/24/98.

\subsection*{APPENDIX}

General method of calculation of high temperature asymptotics has been developed in Ref.
        \cite{Weld}. To outline the procedure used in the present paper let us consider
        the fermion contribution as an example:  $$
f^{(1)}_{r}=\sum^{\infty}_{n=1}(-1)^nK_{|r|}(n\omega). \eqno(A.1)
$$
At first, let us make Mellin's transformation of $K_{|r|}(n\omega)$ with
respect to parameter $n$ \cite{Weld}:  $$
K_{|r|}(s)=\int\limits^{\infty}_{0} K_{|r|}(\omega t)t^{s-1}dt
=\omega^{-s}2^{s-2}  \Gamma\biggl(\frac{s}{2}-\frac{|r|}{2}\biggr)
\Gamma\biggl(\frac{s}{2}+\frac{|r|}{2}\biggr),
$$
where $\Gamma(x)$ is $\Gamma$-function, and substitute it in Eq. (A.1). Then
we find
$$
f_{r}^{(1)}=\frac{1}{8\pi i}\sum^{\infty}_{n=1}(-1)^n
\int\limits^{C_{r+i\infty}}_{C_{r-i\infty}}ds n^{-s}  (\frac{2}{
\omega})^s  \Gamma\biggl(\frac{s}{2}-\frac{|r|}{2}\biggr)
\Gamma\biggl(\frac{s}{2}+\frac{|r|}{2}\biggr),
\eqno(A.2)
$$
where $C_{r} > |r|$ and the integral is calculated along the straight
line parallel to the imaginary axis. If one shifts the integration
variable $s\rightarrow s-r=s^{\prime}$, the integration contour moves
to the right, $ C_{2r}\rightarrow C_{2r}^{\prime}$, for negative $r$,
and to the left, $ C_{r}\rightarrow C_{0}^{\prime}$, in the case of
positive $r$: $2r < C_{2r}^{\prime}< 2r-1; 0 < C_{0}^{\prime}< 1$.
Changing the sequence of the  summation and the  integration in
Eq. (A.2) and taking into account the definition of Riemann's
$\zeta$-function \cite{Abram} $$ \sum^{\infty}_{n-1}\frac{(-1)^n}{n^s} =
(2^{1-s}-1)\zeta(s), $$ one obtains
$$ f_{r}^{(1)}=\frac{1}{8\pi i}
\int\limits^{C_{r+i\infty}}_{C_{r-i\infty}}ds^{\prime}\zeta(s^{\prime})
(2^{1-s^{\prime}}-1)  (\frac{2}{\omega})^{s^{\prime}+r}  \Gamma\biggl(\frac{
s^{\prime}}{2}-\frac{|r|}{2}\biggr)
\Gamma\biggl(\frac{s^{\prime}}{2}+\frac{|r|}{2}\biggr).\eqno(A.3)
$$
It is important to note that the contours $C_{2r}^{\prime},
 C_{0}^{\prime}$ are the same as for $K_{|r|}(x)$ after
shifting of integration variable $s\rightarrow s^{\prime}$. The integral
(A.3) is calculated by closing the contour $ C^{\prime}$ to the left
and summing up the residua of the integrand. For the contour
$C_{2r}^{\prime}$ one should include all the poles of $\Gamma$ and
$\zeta$ -functions. But for positive $r$, $0<C_0<1$, the pole
$s^{\prime}=1$ of $\zeta$-function is out of the contour and should be
excluded.

In the case of fermion contributions the following integrals have to be calculated:
$$
f_{0}^{(1)}=\frac{1}{2}(C+\log(\frac{\omega}{4\pi})) -\frac{2}{\pi}%
\sum^{\infty}_{n=1}(-\frac{\omega^2}{\pi^2})^n  (1-2^{-2n-1}) \frac{%
\Gamma(n+1/2)}{\Gamma(n)}\zeta(2n+1), \eqno(A.4)
$$
$$
f_{r}^{(1)}=-\frac{2^n}{\sqrt{\pi}}(\frac{\omega}{2})^{r-2}
\sum^{\infty}_{n=1}(-\frac{\omega^2}{\pi^2})^n  (1-2^{-2n-2r+1}) \frac{%
\Gamma(n+r-1/2)}{\Gamma(n)}\zeta(2n+2r-1), \eqno(A.5)
$$
$r=1,2...$, C -- Euler constant. As it is seen, in the limit $\omega
\rightarrow 0~ (T \rightarrow \infty)$ the first term in Eq. (A.4) is
dominating. Other terms give the corrections of order $\sim \frac{1}{T},
\frac{1}{T^2}, ...$.

The same procedure has been applied for contributions of boson fields and for particular
        values of $r $ the results of summations are as follows:  $$
\sum^{\infty}_{n=1}\frac{1}{n^2}K_2(n\omega) =
\frac{2\zeta(4)}{\omega^2} -
\frac{\zeta(2)}{2}+\frac{\pi\omega}{6}-\frac{\omega^2}{16}
(\log\frac{4\pi}{ \omega}-C+3/4)+O(\omega^3), \eqno(A.6) $$ $$
\sum^{\infty}_{n=1}\frac{1}{n}K_1(n\omega) = \frac{\zeta(2)}{\omega}
-\frac{\pi}{2}+\frac{\omega}{4}
(\log\frac{4\pi}{\omega}-C+1/2)+O(\omega^2), \eqno(A.7) $$
$$ \sum^{\infty}_{n=1}\frac{1}{n}Y_1(n\omega) = \frac{4\zeta(2)}{\omega}
+\omega(\log\frac{4\pi}{\omega}-C+1/2)+O(\omega^2), \eqno(A.8) $$
$$ \sum^{\infty}_{n=1}K_0(n\omega) =
\frac{1}{2}(C-\log\frac{4\pi}{\omega})+O(\omega). \eqno(A.9)
$$


\begin{thebibliography}{100}

\bibitem{She}    M. Sher,  Phys. Rep. 179, 273 (1989).
\bibitem{Esp}    J.R. Espinosa, Preprint DESY 96-107 IEM-FT-96-133, June 1996.
\bibitem{Oles} P. Olesen, hep-ph/9708320.

\bibitem{Enq}   K. Enqvist, Int. J. Mod. Phys. D7, 331 (1998); astro-ph/9707300.
\bibitem{Vach}  T. Vachaspati, Phys. Lett. B265, 258 (1991).
\bibitem{Shap1}    M. Goivannini and M. Shaposhnikov, Phys. Rev. D57,  2186 (1998).
\bibitem{Shapg}   M. Giovannini and M. E. Shaposhnikov, hep-ph/9708303.
\bibitem{Elm1}  P. Elmfors, K. Enqvist and K. Kainulainen, Phys. Lett. B 440, 269
(1998); hep-ph/9806403.
\bibitem{Shap4}  K. Kajantie, M. Laine, J. Peisa, K. Rummokainen and M. Shaposhnikov,
               Nucl. Phys. B544, 345 (1999); hep-lat/9809004.
\bibitem{Bors1}  V. Skalozub and M. Bordag, Mod. Phys. Lett. A (be published);
hep-ph/9904333.
\bibitem{Bors2}  V. Skalozub and M. Bordag, hep-ph/9905302.
\bibitem{Bors3}  V. Skalozub and V. Demchik, hep-ph/9909550.
\bibitem{Shap2}  V.A. Rubakov and M. E. Shaposhnikov, Uspehi Fiz. Nauk, 166, 493
(1996); hep-ph/9603208 v2 10 Apr 1996. 
\bibitem{Buch}  W. Buchm$\ddot u$ller and O. Philipsen, Phys. Let. B397, 112 (1997).
\bibitem{Kal}   O.K.Kalashnikov, Fortschr. Phys. 32, 325 (1984).
\bibitem{Kirl}  D.A. Kirzhnits and A. D. Linde, Ann. Phys. (NY), 101, 199 (1976).
\bibitem{SVZ1}   A.O.Starinets, A.S. Vshivtsev and V.Ch. Zhukovskii,
                 Phys. Lett. B322, 287 (1994).
\bibitem{SVZ2}   A.S. Vshivtsev, V.Ch. Zhukovskii and A.O. Starinets,
                 Z. Phys. C61, 285 (1994).
\bibitem{Ska3}    V.V. Skalozub, Int. J. Mod. Phys. A 11, 5643 (1996).
\bibitem{Enol}  K. Enqvist and P. Olesen, Phys. Lett. B329 (1994) 195.
\bibitem{Shapj}  M. Joyce and M. Shaposhnikov, astro-ph/9703005.
\bibitem{Baym}  G. Baym, D. B$\ddot o$deker and L. McLerran, Phys. Rev. D53, 662
                (1996).
\bibitem{Cheng} B. Cheng and A.V. Olinto, Phys. Rev. D50, 2451 (1994); G.
               Sigl, A.V. Olinto and K. Jedamzik, ibid. D55, 4582 (1997).
\bibitem{Kibble} T.W. Kibble and A. Viliknkin, Phys. Rev. D52, 679 (1995).
\bibitem{Ahonen} J. Ahonen and K. Enqvist, Phys. Rev. D57, 664 (1998). 
\bibitem{Enqvist98} K. Enqvist, astro-ph/9803196.
\bibitem{Fior}  R. Fiore, A. Tiesi, L. Masperi and A. Megzuand, 
 Mod. Phys. Lett.,A 14, 407 (1999); hep-ph/9804336.
\bibitem{Sch} J. Schwinger, Phys. Rev., 82, 664 (1951).
\bibitem{Ditt1} W. Dittrich and M. Reuter, "Effective Lagrangians in Quantum
Electrodynamics", Lecture Notes in Physics, v. 220, (Springer-Verlag, 1985).
\bibitem{Ska1}  V. V. Skalozub, Sov. J. Nucl. Phys. 28, 113 (1978).
\bibitem{Niol}  N. K. Nielsen and P. Olesen, Nucl. Phys. B144, 376 (1978).
\bibitem{AO1}  J. Ambj$\ddot o$rn and P. Olesen, Int. J. Mod. Phys. A23 (1990) 4525.
\bibitem{Sk2}    V.V. Skalozub, Sov. J. Part. Nucl. 16, 445 (1985). 
\bibitem{Elmp}   P. Elmfors and D. Persson,  Nucl. Phys. B 538, 309 (1999); 
            hep-ph/9806335.
\bibitem{Pers}    D. Persson, hep-ph/9901413.
\bibitem{Amol}     J. Ambj$\ddot o$rn and P. Olesen, Nucl. Phys. B315, 606 (1989);
                 B330 , 193 (1990).
\bibitem{DK}     D. Kirzhnits, JETP Lett. 15, 529 (1972).
\bibitem{ADL}    A.D. Linde, Particle Physics and Inflationary Cosmology
                 (Harwood Academic, New York, 1990).
\bibitem{Cha}    J. Chakrabarti, Phys. Rev. D28 , 2657 (1983); ibd D29 , 1859
                  (1984).
\bibitem{Reu}    M. Reuter and W. Dittrich, Phys. Lett. 144B , 99 (1984).
\bibitem{Fuj}    J. Fujimoto and T. Fukuyama, Z. Phys. C19 , 11 (1983).
\bibitem{Magp}  J.A. Magpantay, W. A. Sayed and C. Mukku, Ann. Phys. 145 (1983) 27.
\bibitem{Sk1}    V.V. Skalozub, Sov. J. Nucl. Phys. 45, 1058 (1987).
\bibitem{Rez}    Yu.Yu. Reznikov and V.V. Skalozub, Sov. J. Nucl. Phys. 46 ,
                 1085 (1987).
\bibitem{Tak}    K. Takahashi,  Z. Phys. C26 , 601 (1985).
\bibitem{Car}    M. Carrington, Phys. Rev. D45 , 2933 (1992).
\bibitem{Gasp}  M. Gasperini, M. Giovannini and G. Veneziano, Phys. Rev. Lett. 75,
       3796 (1995).
\bibitem{Brand}  R.H. Brandenberger, A-C. Davis, A.M. Matheson and M. Trodden, 
        Phys. Lett. B293, 287, (1992).
\bibitem{Migd}  A. B. Migdal, JETP , 62, 1621 (1972).
\bibitem{Gail}  R. M. Gailis, N. E. Frankel and C. P. Dettnann, Phys. Rev. D52, 6901 
(1995).
\bibitem{Abrand}  A. Brandenburg, K. Enqvist and P. Olesen, Phys. Rev. D54, 1291 
(1996).
\bibitem{Sav}  G.K.Savvidy, Phys. Lett. B71, 133 (1977).
\bibitem{Sals}  A. Salam and J. Strathdee, Nucl. Phys. B90, 203 (1975).
\bibitem{SkaW} V. V. Skalozub, Jadernaya Fizika, 35 (1982) 782; 37 (1983) 474.
\bibitem{Din}  M. Dine, R.G. Leigh, P. Huet, A. Linde and D. Linde, Phys. Rev. D46,
        550 (1992).
\bibitem{Kolb}   M. Gleiser and E.W. Kolb, FERMILAB-Pub-92/222-A, NSF-ITP-92-102,
        June 1992.

\bibitem{Cabk}   A. Cabo, O.K. Kalashnikov  and A.E. Shabad, Nucl. Phys. B185, 473 
(1981).
\bibitem{MacD}    S. MacDowell and O. T$\ddot o$rnquist, Phys. Rev. D45,
                 3833 (1992).
\bibitem{Cab}    A. Cabo, Fortschr. Phys. 29, 495 (1981).
\bibitem{Chen}  Ta-Pie Cheng, Ling-Fong  Li, Gauge Theory of Elementary Particle 
        Physics , Clarendon Press-Oxford, 1984.
\bibitem{Schw}  J. Schwinger, Phys. Rev. D7, 1696 (1973).
\bibitem{VZM}    A.S. Vshivtsev, V. Ch. Zhukovskii and B. V. Magnitskii,
                DAN USSR, 314, 175 (1990); A. V. Borisov, A. S. Vshivtsev, V.
            Ch. Zhukovskii and P. A. Eminov, Uspehi Fiz. Nauk, 167, 241 (1997).
\bibitem{Elm}    P. Elmfors, D. Persson and B-S. Skagerstam, Phys. Rev. Lett.
                 71,480 (1993).
\bibitem{Weld} H.E. Haber and A. Weldon, J. Math. Phys. 23, 1852 (1982).
\bibitem{Kap}   J.I.Kapusta, Finite-temperature field theory, Cambridge
 University Press 1989.
\bibitem{SS}  V.Skalozub and A.Strel'chenko, In Proceedings of IY Int. 
Freedman. Conf, 18 - 25 June, 1998, St.-Petrburg, Russia.
\bibitem{Lain}  M. Laine and K. Rummukainen, hep-lat/9809045.
\bibitem{BBS} O.A.Borisenko, J.Bohacik and V.V.Skalozub, Fortschr. Phys. 43, 301 
(1994).
\bibitem{SRV}  V.S. Vanyashin, Yu. Yu. Reznikov and V. V. Skalozub, Soviet    J.
           Nucl. Phys. 33, 227 (1981).
\bibitem{Skav}  V. V. Skalozub and V. S. Vanyashin, Fortschr. Phys. 40, 739 (1992).
\bibitem{Kaja}  K. Kajantie, M. Laine, K. Rummukainen and M. E. Shaposhnikov,
                           Nucl. Phys. B458, 90 (1996);  Nucl. Phys. B466, 189 (1996).
\bibitem{Odin} E. Elizalde, S.D. Odintsov, A. Romeo, A.A. Bytsenko and S. Zerbini,
 Zeta regularization techniques with applications (World Sci., Singapoure, 1994).
\bibitem{Schm}  M. G. Schmidt, Talk at conference  (Heidelberg, July 1998).
\bibitem{AO2}    H.B. Nielsen and M. Ninomiya, Nucl.Phys. B163, 57 (1980);
                 B169, 309 (1980); J. Ambj$\ddot o$rn and P. Olesen, Nucl. Phys.
B170, 265 (1980).
\bibitem{Abram} M. A. Abramovirs and I. A. Stegun, Handbook of Mathematical
        Functions, Dover, New York, 1972.
\end{thebibliography}
\end{document}